\documentclass[12pt,preprint]{aastex}
\begin{document}

\title{The Secular Evolution of a 
  Close Ring--Satellite System:\\
  The Excitation of Spiral Density Waves at a 
  Nearby Gap Edge\vspace*{0.25in}
}

\author{
  Joseph M. Hahn
}
\affil{
  Space Science Institute\\
  10500 Loring Drive\\
  Austin, TX, 78750\\
  email: jhahn@spacescience.org\\
  phone: 512-291-2255\vspace*{0.25in}
}

\author{
  Accepted for publication in the {\it Astrophysical Journal}\\
  \vspace{0.25in}
}

\begin{abstract}

The Lagrange planetary equations are used to study to secular evolution of
a small, eccentric satellite that orbits within a narrow gap in a
broad, self-gravitating planetary ring. These equations show that
the satellite's secular perturbations of the ring will excite
a very long-wavelength spiral density wave that propagates
away from the gap's outer edge. The amplitude of these waves,
as well as their dispersion relation, are derived here. That dispersion relation
reveals that a planetary ring can sustain two types of density waves:
long waves that, in Saturn's A ring, would have wavelengths of 
$\lambda\sim{\cal O}(100)$ km, and short waves that tend to be very
nonlinear and are expected to quickly damp. The excitation
of these waves also transports angular momentum from the ring to the
satellite in a way that
damps the satellite's eccentricity $e$, which also tends to
reduce the amplitude of subsequent waves. The rate of
eccentricity damping due to this wave action is then compared
to the rates at which the satellite's Lindblad and corotation
resonances alter the satellite's $e$. These results are then
applied to the gap-embedded Saturnian satellites Pan
and Daphnis, and the long-term stability of their eccentricities
is assessed.

\end{abstract}


\section{Introduction}
\label{intro}

The following considers the secular gravitational perturbations that
are exerted between a small eccentric satellite and a nearby planetary ring.
This investigation is also a followup to the study described in
\cite{H07}, which examined the secular 
evolution of a small inclined satellite. There it was shown that the
inclined satellite would launch a very long-wavelength spiral bending
wave at the ring's nearby edge. 
Since the excitation of the bending wave also
communicates in-plane angular momentum from the satellite to the
ring, that ring-satellite interaction resulted in a very vigorous damping
of the satellite's inclination.
A related problem was also considered by
\cite{GS03}, who showed that an eccentric 
perturber orbiting in a gap in a pressure-supported gas disk
will launch density waves at the gap's edge having such a
long wavelength that a global standing wave emerges in the disk.
The work described below considers a related but distinct problem, 
that of an eccentric satellite
orbiting in a narrow gap in a self-gravitating planetary ring. 
Here we show that the
secular perturbations from the eccentric
satellite also launches relatively long-wavelength density waves at the gap's
outer edge. The principal goal of this study will be to derive
the amplitude of these waves and their dispersion relation,
which in turn will yield other useful properties, such as the 
wavenumber and the waves' group velocity. Since the mathematics
of this density wave problem is very similar to that already developed
for bending waves, the derivations presented here will be
succinct. However, the reader interested in a more 
verbose description of a similar problem is referred to \cite{H07}.

Another goal of this effort will be to derive the rate at which
the satellite's eccentricity $e$ is altered by the excitation
of these density waves at the gap's edge. This is of interest because
the satellite's $e$ also varies due to its many Lindblad resonances
in the ring, which tend to pump up the satellite's $e$, while
its many corotation resonances in the ring tend to damp
the satellite's $e$ \citep{GT81, GT82}. Although $e$-damping
due to the corotation torque can dominate over $e$-excitation from
the Lindblad torque, the long-term stability of the satellite's eccentricity
is still uncertain since the corotation torque is operative only
when these particular resonances are not saturated. Our purpose here
is to determine whether the secular interaction described below
is also a stabilizing process that leads to
a net damping of the satellite's eccentricity.

The following Section begins with the Lagrange planetary equations,
which are used to derive the amplitude of these waves and their
dispersion relation. Section \ref{e-damping} then uses those results to
determine the rate at which the satellite's eccentricity
evolves in response to the wave it launches at the gap edge. 
Those analytic results are then confirmed in Section \ref{sims}
via a numerical simulation of these waves, with conclusions reported in
Section \ref{summary}.

\section{Equations of motion}
\label{EOM}

Consider a planetary ring that is perturbed by a single satellite, 
with both orbiting an oblate planet. The Lagrange planetary equations
give the rates at which a ring particle's 
orbital eccentricity $e$ and longitude of periapse $\tilde{\omega}$ 
vary with time $t$ due to these perturbations \citep{BC61, MD99}:
\begin{equation}
  \label{edot}
  \dot{e}\simeq-\frac{1}{na^2e}\frac{\partial R}{\partial\tilde{\omega}}
    \quad\mbox{and}\quad
  \dot{\tilde{\omega}}\simeq\frac{1}{na^2e}\frac{\partial R}{\partial e},
\end{equation}
where $R$ is the disturbing function for a small ring particle 
having a semimajor axis
$a$ and mean motion $n\simeq\sqrt{GM/a^3}$, 
where $G$ is the gravitation constant and $M$ is the mass of the 
central planet, and all eccentricities are small, 
$e\ll1$.  The total disturbing
function for a ring particle is 
$R=R_{\mbox{\scriptsize disk}}+R_{\mbox{\scriptsize sat}}+
R_{\mbox{\scriptsize obl}}$,
where the three terms account for the gravitational perturbations
that are due to the ring's gravity (which is
treated here as a broad disk), the satellite's perturbations, 
and that due to the planet's oblate figure. The particle's
equation of motion is thus the sum of three parts:
\begin{equation}
  \dot{e}=\left.\dot{e}\right|_{\mbox{\scriptsize disk}}+
    \left.\dot{e}\right|_{\mbox{\scriptsize sat}}
    \quad\mbox{and}\quad
    \dot{\tilde{\omega}}=
    \left.\dot{\tilde{\omega}}\right|_{\mbox{\scriptsize disk}}+
    \left.\dot{\tilde{\omega}}\right|_{\mbox{\scriptsize sat}}+
    \left.\dot{\tilde{\omega}}\right|_{\mbox{\scriptsize obl}},
\end{equation}
noting that oblateness does not alter eccentricities. 
And because we are only dealing with the system's secular 
perturbations, the system's semimajor axes $a$ are all constant
\citep{BC61}.

The amplitude of a spiral density wave that is in a 
steady--state does not vary with time,
so the disk eccentricities obey
$\dot{e}(a)=0$ throughout the disk. A persistent spiral pattern must also
rotate with a constant angular velocity $\omega$, so
\begin{mathletters}
  \label{ss}
  \begin{eqnarray}
    \label{edot_ss}
    \left.\dot{e}\right|_{\mbox{\scriptsize disk}}&=&
    -\left.\dot{e}\right|_{\mbox{\scriptsize sat}}\\
    \label{omega_dot_ss}
    \mbox{and}\quad\omega&=&
      \left.\dot{\tilde{\omega}}\right|_{\mbox{\scriptsize disk}}+
      \left.\dot{\tilde{\omega}}\right|_{\mbox{\scriptsize sat}}+
      \left.\dot{\tilde{\omega}}\right|_{\mbox{\scriptsize obl}}
      =\mbox{ constant}.
  \end{eqnarray}
\end{mathletters}
These equations are used to obtain the wave amplitude $e(a)$ 
throughout the disk, and the waves' dispersion relation 
$\omega(k)$ as a function of the wavenumber $k$.

\subsection{wave amplitude}
\label{amplitude}

First calculate the rate at which
the planetary ring perturbs itself.
This ring is treated as a broad disk composed 
of many narrow, concentric annuli that have a
mass $\delta m(a)$, eccentricity $e(a)$, and longitude
of periapse $\tilde{\omega}(a)$ that are regarded as functions
of the rings' semimajor axes $a$. Each annulus also has
a half-thickness $h$ that is due to the ring
particles' dispersion velocity. Now 
suppose that the annulus at $a$ is perturbed by another annulus of mass 
$\delta m'$ and radius $a'$; the perturbed annulus will then have a
disturbing function
\begin{equation}
  \label{deltaR0}
  \delta R=\frac{G\delta m'}{4a}
    \left[\frac{1}{2}f(\alpha)e^2
    +g(\alpha)ee'\cos(\tilde{\omega}-\tilde{\omega}')  \right],
\end{equation}
where $a, e, \tilde{\omega}$ are the orbit elements of the 
perturbed annulus, the primed quantities refer to the 
perturbing annulus, and the $f(\alpha)$ and $g(\alpha)$ 
functions are
\begin{mathletters}
  \label{fg}
  \begin{eqnarray}
    \label{f(a)}
    f(\alpha)&=&\alpha\tilde{b}^{(1)}_{3/2}(\alpha)
      -6\mathfrak{h}^2\alpha^2\tilde{b}^{(0)}_{5/2}(\alpha)\\
    \label{g(a)}
    \mbox{and}\quad
    g(\alpha)&=&-\alpha\tilde{b}^{(2)}_{3/2}(\alpha)
      +6\mathfrak{h}^2\alpha^2\tilde{b}^{(1)}_{5/2}(\alpha),
  \end{eqnarray}
\end{mathletters}
which depend on the semimajor axis ratio $\alpha=a'/a$ and
the disk's dimensionless scale height $\mathfrak{h}=h/a\ll1$,
which is presumed small. This disturbing function is
derived in \cite{H03}, and it differs somewhat from
the more familiar disturbing function for a point-mass perturber
due to the annuli's finite thickness $h$, which also
softens the Laplace coefficients that appear in $\delta R$. Those
softened Laplace coefficients  are
\begin{equation}
  \label{lc}
  \tilde{b}^{(m)}_{s}(\alpha)=\frac{2}{\pi}\int_0^\pi
    \frac{\cos(m\varphi)d\varphi}
    {[(1+\alpha^2)(1+\mathfrak{h}^2)-2\alpha\cos\varphi]^{s}}.
\end{equation}
Note that when the annuli are infinitesimally thin, 
$\mathfrak{h}\rightarrow0$, and the 
softened Laplace coefficients 
$\tilde{b}^{(m)}_{s}(\alpha)$ are equivalent to the conventional
Laplace coefficients $b^{(m)}_{s}(\alpha)$, for which
$f(\alpha)\rightarrow\alpha b^{(1)}_{3/2}(\alpha)$ and
$g(\alpha)\rightarrow-\alpha b^{(2)}_{3/2}(\alpha)$,
and the disturbing function $\delta R$ becomes equivalent 
to that due to a point--perturber of mass 
$\delta m'$ (e.g., \citealt{BC61}).

The perturbing ring's mass is
$\delta m'=2\pi\sigma'a'da'$ where $\sigma'=\sigma(a')$ 
is the mass surface density of the perturbing annulus of radius $a'$
and radial width $da'$. The perturbing ring's 
semimajor axis is written $a'=a(1+x')$,
where $x'=(a'-a)/a=\alpha-1$ is the fractional distance 
between the perturbing ring $a'$ and
the perturbed ring $a$. Also define the 
normalized disk mass as $\mu_d(a)\equiv\pi\sigma a^2/M$,
so that the factor $G\delta m'/4a$ in Eqn.\ (\ref{deltaR0})
becomes $\mu_d'(na)^2dx'/2\alpha$ where $dx'=da'/a$
is the perturbing ring's fractional width. The
disturbing function for ring $a$
due to perturbations from ring $a'$, Eqn.\ (\ref{deltaR0}), 
then becomes
\begin{equation}
  \label{deltaR}
  \delta R=\frac{1}{2}\mu_d'(na)^2\alpha^{-1}
    \left[\frac{1}{2}f(x')e^2
    +g(x')ee'\cos(\tilde{\omega}-\tilde{\omega}')\right]dx'
\end{equation}
where $f(x')$ and $g(x')$ is shorthand for Eqns.\ (\ref{fg}) 
evaluated at $\alpha=1+x'$. Inserting this into Eqn.\ (\ref{edot})
then shows that ring $a'$ alters the 
eccentricity of ring $a$ at the rate
\begin{equation}
  \label{delta-dot-e}
  \delta\dot{e}=-\frac{1}{na^2e}
    \frac{\partial(\delta R)}{\partial\tilde{\omega}}
    =\frac{1}{2}\mu_d'n\alpha^{-1}g(x')e'
    \sin(\tilde{\omega}-\tilde{\omega}')dx'.
\end{equation}

\subsubsection{ring--disk evolution}
\label{ring_disk evol}

\begin{figure}
\epsscale{1.0}
\plotone{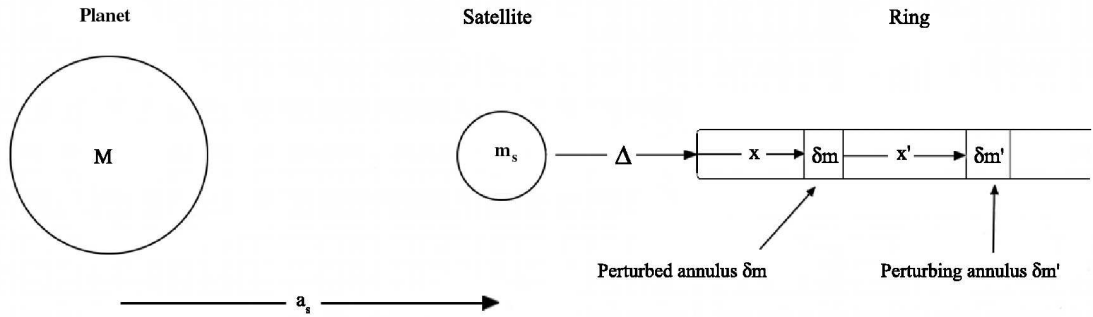}
\figcaption{
  \label{geometry}
  Schematic showing the geometry of the ring--satellite system
  seen edge-on. A satellite of mass $m_s$ and semimajor axis $a_s$
  orbits interior to a broad planetary ring that extends to infinity.
  The satellite's distance from the ring's inner edge is $\Delta$ 
  in units of the satellite's semimajor axis $a_s$. 
  A perturbed annulus in the ring has mass $\delta m$,
  and it lies a fractional distance $x$ away from the ring's inner edge,
  while the perturbing ring has mass $\delta m'$ 
  and lies a fractional distance $x'$ from the perturbed ring.
}
\end{figure}

The total rate at which the
entire disk alters the eccentricity of ring $a$ is the above 
with $x'$ integrated across the disk, so 
$\left.\dot{e}\right|_{\mbox{\scriptsize disk}}=
\int_{\mbox{\scriptsize disk}}\delta\dot{e}$.
For the moment, consider a one-sided disk, one that orbits wholly
exterior to the satellite, where
$\Delta$ is the fractional distance between the satellite's orbit
and the disk's inner edge. The geometry is sketched in Fig.\ \ref{geometry},
which shows that the integration variable $x'$ then ranges from $-x$ to
$+\infty$, so the ring's eccentricity varies at the rate
\begin{equation}
  \label{edot_disk0}
  \left.\dot{e}\right|_{\mbox{\scriptsize disk}}=
    \int_{\mbox{\scriptsize disk}}\delta\dot{e}=
    \frac{1}{2}n\int_{-x}^{\infty}
    \mu_d'(x')\alpha^{-1}g(x')e'\sin(\tilde{\omega}-\tilde{\omega'})dx'.
\end{equation}
This integral will be dominated by the contributions from
nearby annuli that lie a small distance $x'$ away. 
In the $|x'|\ll1$, $\mathfrak{h}\ll1$ limit, 
the softened Laplace coefficients that appear in the $g(\alpha)$
function, Eqn.\ (\ref{g(a)}), are
\begin{mathletters}
  \label{b_approx}
    \begin{eqnarray}
      \label{b1_3haf}
      \tilde{b}^{(m)}_{3/2}(x')&\simeq&\frac{2}{\pi(x'^2+2\mathfrak{h}^2)}\\
      \label{b1_5haf}
      \mbox{and}\quad\tilde{b}^{(m)}_{5/2}(x')&\simeq&
        \frac{4}{3\pi(x'^2+2\mathfrak{h}^2)^2}
  \end{eqnarray}
\end{mathletters}
\citep{H03}, so $\alpha^{-1}g(\alpha)$ in Eqn.\ (\ref{edot_disk0})
becomes
\begin{equation}
  \alpha^{-1}g(\alpha)\simeq-\frac{2}{\pi}
    \frac{x'^2-2\mathfrak{h}^2}{(x'^2+2\mathfrak{h}^2)^2}
\end{equation}
in this approximation. Due to the steep dependence of 
$g(\alpha)$ on $x'$, we can replace the eccentricity $e'(x')$ 
and disk mass $\mu_d'(x')$ in Eqn.\ (\ref{edot_disk0}) with their values
evaluated at the perturbed ring, which lies at $x'=0$, so
$e'\simeq e$ and $\mu_d'\simeq\mu_d=\pi\sigma a^2/M$, which
are then pulled out of the integral so that
\begin{equation}
  \label{edisk0}
  \left.\dot{e}\right|_{\mbox{\scriptsize disk}}\simeq-\frac{1}{\pi}
    \mu_den\int^\infty_{-x}
    \frac{x'^2-2\mathfrak{h}^2}{(x'^2+2\mathfrak{h}^2)^2}
    \sin(\tilde{\omega}-\tilde{\omega}')dx'.
\end{equation}

A spiral wave will have a wavelength $\lambda\simeq2\pi/|k|$,
where $k=-\partial\tilde{\omega}/\partial a$ 
is the wavenumber of the spiral density wave. Thus the 
$\tilde{\omega}-\tilde{\omega}'$ in Eqn.\ (\ref{edisk0})
can also be written as
\begin{equation}
  \label{wavenumber_exact}
  \tilde{\omega}-\tilde{\omega}'(a')=-\int_{a'}^a k(r)dr.
\end{equation}
However, most of the contribution to the integral in Eqn.\ (\ref{edisk0})
will be due to nearby annuli that lie
a wavelength $\lambda$ away.
But if $k$ also varies slowly with $a$, then
it can be treated as a constant over that wavelength,
so Eqn.\ (\ref{wavenumber_exact}) is
$\tilde{\omega}-\tilde{\omega'}\simeq-k(a-a')=kax'$, and
Eqn.\ (\ref{edisk0}) then becomes
\begin{equation}
  \label{edisk1}
  \left.\dot{e}\right|_{\mbox{\scriptsize disk}}\simeq-\frac{1}{\pi}
    ka\mu_den\int^\infty_{-|k|ax}
    \frac{y^2-H^2}{(y^2+2H^2)^2}\sin(y)dy
    \equiv-\frac{1}{\pi}A'_H(|k|ax)ka\mu_den
\end{equation}
after replacing the $x'$ integration variable with $y=|k|ax'$,
which is the distance from the ring--edge in units of
$2\pi$ wavelengths, and replacing the disk's scale height 
$\mathfrak{h}$ with $H\equiv\sqrt{2}\mathfrak{h}|k|a$, 
which is roughly the disk's vertical thickness in wavelength units.
In the above, the function $A'(z)$ is
\begin{equation}
  \label{A}
  A'_H(z)=\int_z^\infty\frac{y^2-H^2}{(y^2+H^2)^2}\sin ydy,
\end{equation}
noting that Eqn.\ (\ref{edisk1}) is also odd in $y$. 
We will be interested in a disk whose vertical thickness is
small compared to the wavelength, so $H\ll1$, and
$A'_H(z)\simeq\sin(z)/z-\mbox{Ci}(z)$, 
where $\mbox{Ci}(z)$ is the cosine integral
of \citet{AS72}.  Far downstream, where $z\gg1$,
$\mbox{Ci}(z)\simeq\sin(z)/z-\cos(z)/z^2+\cal{O}$$(z^{-3})$ 
\citep{AS72}, so 
\begin{equation}
  \label{A_downstream}
  A'_H(z)\simeq\frac{\cos(z)}{z^2}
\end{equation}
downstream.  Evidently, $A'_H(z)$ is very similar to
the $A_H(z)$ function of \cite{H07}, 
which is also plotted in Fig.\ 2 there.

Also keep in mind that Eqn.\ (\ref{edisk1}) was derived with the
understanding that the wavenumber $k$ varies little over
a single wavelength.  Section \ref{wavenumber} will quantify when that
assumption breaks down.

\subsubsection{ring--satellite evolution, and the wave amplitude}
\label{wave-amplitude}

The ring at semimajor axis $a$ is also perturbed by the satellite, 
and the disturbing function $R_s$ due to the satellite is 
Eqn.\ (\ref{deltaR0}) with $\delta m'$ replaced by the satellite's 
mass $m_s$, which can be written
\begin{equation}
  \label{Rs}
  R_s=\frac{1}{4}\mu_s(na)^2
    \left[\frac{1}{2}f(\alpha)e^2+
    g(\alpha)ee_s\cos(\tilde{\omega}-\tilde{\omega}_s)\right],
\end{equation}
where $\mu_s=m_s/M$ is the satellite's mass in units of the 
central planet's mass, and
$\alpha=a_s/a=(1+\Delta+x)^{-1}\simeq1-(\Delta+x)$. The 
satellite's perturbation causes the ring's eccentricity 
to vary at the rate
\begin{equation}
  \label{edot_sat0}
  \left.\dot{e}\right|_{\mbox{\scriptsize sat}}=
    -\frac{1}{na^2e}\frac{\partial R_s}{\partial\tilde{\omega}}
    =\frac{1}{4}\mu_s ng(\alpha)e_s
    \sin(\tilde{\omega}-\tilde{\omega}_s).
\end{equation}
The perturbed ring lies a fractional distance $x+\Delta$ away 
from the satellite (see Fig.\ \ref{geometry}), with both well separated
so that $\Delta\gg\mathfrak{h}$. Since the ring also lies
in the wave--excitation zone near the satellite, 
\begin{equation}
  g(\alpha)\simeq-\alpha\tilde{b}^{(2)}_{3/2}(\alpha)
    \simeq-\frac{2}{\pi(x+\Delta)^2}
\end{equation}
by Eqns.\ (\ref{g(a)}) and (\ref{b1_3haf}).
Also write the longitude difference in Eqn.\ (\ref{edot_sat0}) as 
$\tilde{\omega}-\tilde{\omega}_s\simeq-kax+\phi_o$,
where the angle $\phi_o$ allows for the possibility that 
the annulus nearest the satellite at $x=0$ has a longitude of 
periapse that differs from the satellite's longitude
$\tilde{\omega}_s$ by angle $\phi_o$. Thus
\begin{equation}
  \label{edot_sat}
  \left.\dot{e}\right|_{\mbox{\scriptsize sat}}\simeq
    \frac{\mu_se_sn}{2\pi(x+\Delta)^2}\sin(kax-\phi_o)
\end{equation}
is the rate at which the satellite alters the ring's eccentricity.

When the wave is in steady--state, the two $e$--excitation rates,
Eqns.\ (\ref{edisk1}) and (\ref{edot_sat}), are balanced,
which provides the amplitude of the density wave:
\begin{equation}
  \label{e_exact}
  \frac{e(z)}{e_s}=\frac{|k|a}{2}\frac{\mu_s}{\mu_d}
    \frac{\sin(z-s_k\phi_o)}{(z+|k|a\Delta)^2A'_H(z)}
\end{equation}
where $s_k=\mbox{sgn}(k)$ 
and $z=|k|ax$ is the downstream distance in units of
$2\pi$ wavelengths. Now recall that
$A'_H(z)\simeq\cos(z)/z^2$ far downstream
where $z\gg1$ (Eqn.\ \ref{A_downstream}). Since
the downstream wave amplitude $e(z)$ should be a finite constant
for all $z\gg1$, this then tells us that
the longitude offset is $\phi_o=\pm\pi/2$,
so $e/e_s\simeq -(|k|a/2)(\mu_s/\mu_d)s_k\sin\phi_o$. 
These eccentricities must also be positive, so 
$\sin\phi_o=\pm1=-s_k$, and the density wave amplitude then becomes
\begin{equation}
  \label{edownstream}
  \frac{e}{e_s}\simeq\frac{|k_0|a\mu_s}{2\mu_d},
\end{equation}
where $|k_0|$ is the initial wavenumber at $x=0$, where the wave
is excited at the disk's inner edge. An identical expression
was also obtained in \cite{H07} for the amplitude of the
bending wave that an inclined satellite would launch at the ring's
edge. 

\subsection{dispersion relation}
\label{DR}

Further use of the wave amplitude, Eqn.\ (\ref{edownstream}), 
requires knowing the initial wavenumber $k_0$, which is obtained
from the waves' dispersion relation, Eqn.\ (\ref{omega_dot_ss}).
The first term in that equation, 
$ \left.\dot{\tilde{\omega}}\right|_{\mbox{\scriptsize disk}}$, is the rate
at which the disk drives its own precession. 
The rate that a single annulus at $a$ precesses due to the
secular perturbations from another annulus at $a'$ is
\begin{equation}
  \label{delta_dot_omega}
  \delta\dot{\tilde{\omega}}=
    \frac{1}{na^2e}\frac{\partial(\delta R)}{\partial e}
    =\frac{1}{2}\mu_d'n\alpha^{-1}
    \left[f(x')+g(x')\frac{e'(x')}{e}
    \cos(\tilde{\omega}-\tilde{\omega}')\right]dx'
\end{equation}
where $\delta R$ is Eqn.\ (\ref{deltaR}). 
The total precession rate due to the disk's self--gravity is
$\left.\dot{\tilde{\omega}}\right|_{\mbox{\scriptsize disk}}=
\int_{\mbox{\scriptsize disk}}\delta \dot{\tilde{\omega}}$,
where the integration proceeds across the entire disk. Again, the integrand
is a steep function of $x'$ due to the softened Laplace coefficients
that are present in the $f$ and $g$ functions, which allows us
to replace the quantities $e'(x')$ and $\mu_d'(x')$ 
with the constants $e$ and $\mu_d$ and to pull them out of the integral.
And since $f(x')\simeq-g(x')$ when $|x'|\ll1$ 
(from Eqns.\ \ref{fg} and \ref{b_approx}), 
\begin{mathletters}
  \begin{eqnarray}
    \label{disk_precess0}
    \left.\dot{\tilde{\omega}}\right|_{\mbox{\scriptsize disk}}&\simeq&
      \frac{1}{\pi}\mu_dn\int^{\infty}_{-x}
      \frac{x'^2-2\mathfrak{h}^2}{(x'^2-2\mathfrak{h}^2)^2}
      [1-\cos(|k|ax')]dx'\\
      &=&\frac{2}{\pi}\mu_dn\int^{\infty}_{-x}
      \frac{x'^2-2\mathfrak{h}^2}{(x'^2-2\mathfrak{h}^2)^2}
      \sin^2(|k|ax'/2)dx'\\
    \label{disk_precess}
    &\equiv&B'_H(|k|ax)|k|a\mu_dn
  \end{eqnarray}
\end{mathletters}
where
\begin{equation}
  \label{Bexact}
  B'_H(z)=\frac{2}{\pi}\int_{-z}^\infty
    \frac{y^2-H^2}{(y^2+H^2)^2}\sin^2(y/2)dy.
\end{equation}
When the disk is much thinner than the wavelength, 
$H=\sqrt{2}\mathfrak{h}|k|a\ll1$, and 
\begin{equation}
  B_H'(z)\simeq\frac{1}{2}+\frac{1}{\pi}\mbox{Si}(z)+\frac{\cos z-1}{\pi z}
\end{equation}
where $\mbox{Si}(z)$ is the sine integral of \citet{AS72}.
In this limit the $B'_H(z)$ function is 
identical to the $B_H(z)$ function of \cite{H07}, with both
taking values of $1/2<B'_H<1$.
And far downstream, where $z\rightarrow\infty$, the $B'_H$
integral evaluates to
\begin{equation}
  \label{B_infinity}
  B_H^{'\infty}\equiv \lim_{z\rightarrow\infty}B'_H(z)=e^{-H}
\end{equation}
for any value of $H$, which, by the way, does differ a bit from the $B_H$ 
function of \cite{H07}.
The function $B'_H(z)$ is a dimensionless measure of the rate at which the 
disk drives its own precession, and this quantity 
becomes small when $H\gtrsim1$,
or when the disk thickness exceeds the
wavelength. Evidently, a thick disk having $H\gg1$ is less
likely to sustain a density wave.

The satellite's gravity also precesses the planetary ring,
at the rate
\begin{eqnarray}
 \left.\dot{\tilde{\omega}}\right|_{\mbox{\scriptsize sat}}&=&
    \frac{1}{na^2e}\frac{\partial R_s}{\partial e}
    =\frac{1}{4}\mu_sn
    \left[f(\alpha)+g(\alpha)\frac{e_s}{e}
    \cos(\tilde{\omega}-\tilde{\omega_s})\right]\\
  \label{sat_precess}
  &\simeq&\left[\frac{\mu_s}{2\pi(x+\Delta)^2} 
    +\frac{\mu_d\sin(|k|ax)}{\pi|k|a(x+\Delta)^2}\right]n
\end{eqnarray}
when $e_s/e$ is replaced with Eqn.\ (\ref{edownstream}).
The disturbing function associated with the central 
planet's oblateness is also
\begin{equation}
  R_{\mbox{\scriptsize obl}}\simeq
    \frac{3}{4}J_2e^2\left(\frac{R_p}{a}\right)^2(an)^2,
\end{equation}
where $J_2$ is the planet's second zonal harmonic 
and $R_p$ is the planet's radius \citep{MD99}, so
precession due to oblateness is 
\begin{eqnarray}
  \label{oblate_exact}
  \left.\dot{\tilde{\omega}}\right|_{\mbox{\scriptsize obl}}&=&
    \frac{1}{na^2e}\frac{\partial R_{\mbox{\scriptsize obl}}}{\partial e}
    =\frac{3}{2}J_2\left(\frac{R_p}{a}\right)^2n
  \simeq\left[1-\frac{7}{2}(x+\Delta)\right]
    \left.\dot{\tilde{\omega}}_s\right|_{\mbox{\scriptsize obl}}
\end{eqnarray}
where 
$\left.\dot{\tilde{\omega}}_s\right|_{\mbox{\scriptsize obl}}
\equiv(3J_2/2)(R_p/a_s)^2n_s$
is the rate at which the satellite's orbit precesses due to oblateness,
with $n_s$ being the satellite's mean motion.
Summing Eqns.\ (\ref{disk_precess}), (\ref{sat_precess}), and 
(\ref{oblate_exact}) then
provides the dispersion relation for these spiral density waves,
\begin{equation}
  \label{omega(k)}
  \omega(|k|)\simeq D'(z)\mu_d|k|an
    +\frac{\mu_sn}{2\pi(x+\Delta)^2}
    +\left[1-\frac{7}{2}(x+\Delta)\right]
    \left.\dot{\tilde{\omega}}_s\right|_{\mbox{\scriptsize obl}},
\end{equation}
which is the angular rate at which the spiral density pattern
rotates. The $D'$ function in the above is
\begin{equation}
  \label{D}
  D'(z)=B'_H(z)+\frac{\sin z}{\pi(z+|k|a\Delta)^2},
\end{equation}
where $D'(z)$ is again numerically
similar to the $D(z)$ function of \cite{H07}, which is
also plotted in Fig.\ 2 there, which 
shows that $D'(z)$ takes numerical values of
$1/2\le D(z)\le1$ when $H\ll1$.

\subsubsection{long and short density waves}

When these density waves have propagated far downstream of the
satellite, the middle term in Eqn.\ (\ref{omega(k)}), 
which is the rate at which the satellite
precesses the disk, becomes even more negligible with distance
from the satellite, and the $D'(z)$ function becomes 
$D'\simeq B_H^{'\infty}=e^{-H}$ 
where $H=\sqrt{2}\mathfrak{h}|k|a$ should now be regarded as a 
dimensionless wavenumber. Thus
the downstream dispersion relation is
\begin{equation}
  \label{down_disp_rel}
  \omega(|k|)\simeq \mu_d|k|ane^{-\sqrt{2}\mathfrak{h}|k|a}+
    \left.\dot{\tilde{\omega}}\right|_{\mbox{\scriptsize obl}}(x).
\end{equation}
This dispersion relation can then be expressed 
in a more convenient dimensionless form via the combination
\begin{equation}
  \label{dim_disp_rel}
  \omega^\star\equiv\frac{\sqrt{2}\mathfrak{h}}{\mu_dn}
    [\omega(|k|)-\left.\dot{\tilde{\omega}}\right|_{\mbox{\scriptsize obl}}]
    =He^{-H}.
\end{equation}

\begin{figure}
\epsscale{1.0}
\plotone{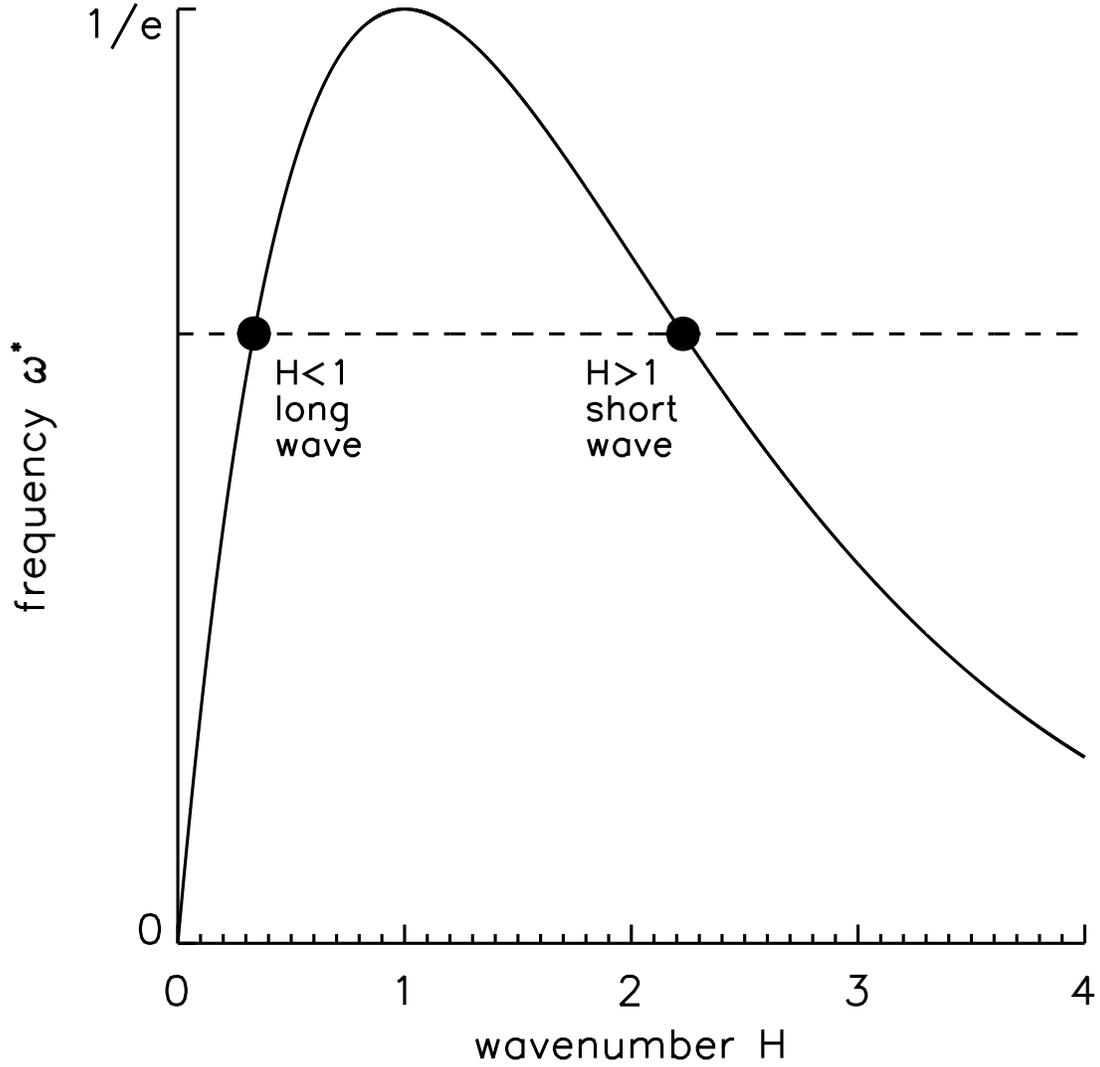}
\figcaption{
  \label{disp_fig}
  The dimensionless dispersion relation, Eqn.\ (\ref{dim_disp_rel}),
  plotted versus the dimensionless wavenumber $H=\sqrt{2}\mathfrak{h}|k|a$.
  Note that $\omega^\star$ has a maximum value of 
  $\omega^\star_{\mbox{\scriptsize max}}=\exp(-1)\simeq0.368$,
  which occurs when the wavenumber $H=1$.
  When the spiral density pattern precesses at an angular
  rate $\omega^\star<\omega^\star_{\mbox{\scriptsize max}}$,
  this dispersion relation has two solution: a long-wavelength
  wave that has a wavenumber $H<1$, and a short-wavelength
  wave that has wavenumber $H>1$.
}
\end{figure}

Figure \ref{disp_fig} plots this
dimensionless dispersion relation $\omega^\star$
versus the dimensionless wavenumber $H$, which shows that 
$\omega^\star$ has a maximum value of 
$\omega^\star_{\mbox{\scriptsize max}}=\exp(-1)\simeq0.368$,
which occurs for a wavenumber $H=1$.
This figure also shows that as long as the density wave's spiral
pattern rotates at an angular rate 
$\omega^\star<\omega^\star_{\mbox{\scriptsize max}}$,
then the disk can sustain two types of waves: long waves
that have a wavenumber $H<1$, and short waves
that have $H>1$, noting that these waves are called such since 
their wavelength is $\lambda=2\pi/|k|=2\sqrt{2}\pi h/H$.
This upper limit on $\omega^\star$ also tells us that the disk
can sustain these density waves when the spiral density
pattern does not rotate too fast, namely, when
\begin{equation}
  \omega<\frac{\omega^\star_{\mbox{\scriptsize max}}\mu_dn}{\sqrt{2}\mathfrak{h}}+
    \left.\dot{\tilde{\omega}}\right|_{\mbox{\scriptsize obl}}
\end{equation}
(see also \citealt{H03}).

\subsubsection{group velocity}
\label{group velocity}

The waves' group velocity is 
$c_g=\partial\omega/\partial k$ \citep{T69, S84, BT87}, which becomes
\begin{equation}
  \label{cg}
  c_g=s_k\frac{\partial\omega}{\partial |k|}=s_k(1-H)e^{-H}\mu_dan
\end{equation}
when the downstream dispersion relation is differentiated 
(Eqn.\ \ref{down_disp_rel}); this is the rate of the spiral wave's
radial propagation. 
Since the satellite is launching 
outward--propagating density waves from the disk's inner edge, 
$c_g>0$, so Eqn.\ (\ref{cg})  implies that the 
satellite can launch long $H<1$ waves that have
$s_k=\mbox{sgn}(k)=1$, or short $H>1$ waves that have 
$s_k=-1$. Spiral density waves having $k>0$
are called trailing waves, since the more distant parts
of a spiral arm trail in longitude, while leading waves
have $k<0$. 

Section \ref{assumptions} will show that any waves
excited by the Saturnian satellites Pan and Daphnis,
both of which inhabit narrow gaps in Saturn's main A ring,
would have a dimensionless wavenumber $H\ll1$.
This means that these satellites could launch
long trailing waves that have $s_k=1$ 
that would propagate outwards at a rate $c_g\simeq\mu_dan$
(e.g., \citealt{H03}).
And since $\sin\phi_o=-s_k=-1$ (see Section \ref{wave-amplitude}), then 
$\phi_o=-\pi/2$, which means that the longitude of periapse at the 
disk's inner edge would trail the satellite's longitude by $90^\circ$.

Later, Eqn.\ (\ref{k}) will also show that the wavenumber $|k|$ and
hence $H$ will increase as the wave propagates downstream.
So it is possible that a wave might 
travel far enough for the wavenumber $H$ to increase beyond unity. 
If that happens, then
Eqn.\ (\ref{cg}) suggests three possible outcomes:
the trailing wave can continue to advance further downstream
with $c_g>0$ as a short ($H>1$) leading ($s_k=-1$) wave,
or it can reflect ($c_g<0$) as a short ($H>1$) trailing ($s_k=+1$) wave,
or it might do both by spawning both types of wavetrains.
Note, though, that the short wave would propagate at a
substantially slower rate, $|c_g|\sim He^{-H}\mu_dan$,
since $H>1$. This site in the disk where $H=1$ and
$c_g$ changes sign is a turning point for long waves. 
This site is also known as the $Q$--barrier,
since its location depends upon the value of Toomre's
stability parameter $Q\simeq v_dn/\pi G\sigma=\mathfrak{h}/\mu_d$,
where $v_d=hn$ is the particle disk's dispersion velocity
\citep{T64}.

\subsubsection{wavenumber and wavelength}
\label{wavenumber}

The wavenumber $k$ is obtained after calculating 
the satellite's precession rate $\dot{\tilde{\omega}}_s$. 
When the system is in steady state, both the satellite
and the spiral wave precess at the same rate, 
$\dot{\tilde{\omega}}_s=\omega(|k|)$, which then provides a 
simple relation for $k$ that depends only on the system's 
physical constants, namely, $\mu_d$, $\Delta$, and $J_2$.

The satellite's longitude of periapse $\tilde{\omega}_s$ 
precesses due to perturbations from the disk and the 
central planet oblateness, with this precession occurring at the rate
\begin{equation}
  \dot{\tilde{\omega}}_s=
    \left.\dot{\tilde{\omega}}_s\right|_{\mbox{\scriptsize disk}}
    + \left.\dot{\tilde{\omega}}_s\right|_{\mbox{\scriptsize obl}}
\end{equation}
where $\left.\dot{\tilde{\omega}}_s\right|_{\mbox{\scriptsize disk}}=
\int_{\mbox{\scriptsize disk}}\delta\dot{\tilde{\omega}}_s$
is the satellite's precession rate due to the entire disk, where
\begin{equation}
  \delta\dot{\tilde{\omega}}_s=
    -\frac{1}{2}\mu_d'n_s\frac{g(\alpha)}{\alpha}
    \left[1-\frac{e'}{e_s}\cos(\tilde{\omega}_s-\tilde{\omega}')\right]dx'
\end{equation}
is the satellite's precession rate due to a disk annulus of 
radius $a'$ and mass $\delta m'$. 
This is obtained from Eqn.\ (\ref{delta_dot_omega}) when
$n, a, e, \tilde{\omega}$ is replaced with
$n_s, a_s, e_s, \tilde{\omega}_s$, and  the separation 
$x'\rightarrow x'+\Delta$. The satellite's precession rate
due to the entire disk is
\begin{mathletters}
\begin{eqnarray}
  \left.\dot{\tilde{\omega}}_s\right|_{\mbox{\scriptsize disk}}&=&
    \int_{\mbox{\scriptsize disk}}\delta\dot{\tilde{\omega}}_s\simeq
    \frac{1}{\pi}\mu_dn_s\int_0^\infty(x'+\Delta)^{-2}
    \left[1-\frac{e}{e_s}\cos(kax'-\phi_o)\right]dx'\\
  \label{dot_omega_s0}
  &\simeq&\frac{\mu_dn_s}{\pi\Delta}+
    \frac{\mu_sn_s}{2\pi\Delta^2}S(|k_0|a\Delta)
\end{eqnarray}
\end{mathletters}
where
\begin{equation}
  \label{S}
  S(|k_0|a\Delta)\equiv 
    |k_0a\Delta|^2\int_0^\infty\frac{\sin(y)dy}{(y+|k_0|a\Delta)^2}
\end{equation}
is another dimensionless function that depends on the the initial
wavenumber $|k_0|$ and the gap width $\Delta$. This quantity
is also plotted in  Fig.\ 2 of \cite{H07}, which shows that 
$S$ takes numerical values of $0\le S(|k_0|a\Delta)\le 1$.

The first term in Eqn.\ (\ref{dot_omega_s0}) 
is the rate at which the undisturbed
disk precesses the satellite's orbit, while the second term 
is the additional precession that the satellite experiences
due to the density wave in the disk. Note, though, that if the 
satellite instead orbited at the center of a narrow gap in the disk,
then the first term in Eqn.\ (\ref{dot_omega_s0})
would be doubled due to the disk matter orbiting
interior to the satellite. One would also expect 
additional precession to occur due to any density waves
that might be launched in this interior disk. 
However, it will be shown below that
this contribution is unimportant. With this in mind, 
Eqn.\  (\ref{dot_omega_s0}) is generalized
to account for a possible inner disk by writing
\begin{equation}
    \label{dot_omega_s}
  \left.\dot{\tilde{\omega}}_s\right|_{\mbox{\scriptsize disk}}\simeq
    \frac{\varepsilon\mu_dn_s}{\pi\Delta}
    +\frac{\mu_sn_s}{2\pi\Delta^2}S(|k_0|a\Delta)
\end{equation}
where $\varepsilon=1$ if the disk lies entirely interior or exterior to the
satellite, and $\varepsilon=2$ if the satellite instead orbits in the center
of a gap whose fractional half--width is $\Delta$.
The satellite's total precession rate then becomes
\begin{equation}
  \label{omega_s_dot}
  \dot{\tilde{\omega}}_s= 
    \left.\dot{\tilde{\omega}}_s\right|_{\mbox{\scriptsize disk}}+
    \left.\dot{\tilde{\omega}}_s\right|_{\mbox{\scriptsize obl}}=
    \frac{\varepsilon\mu_dn_s}{\pi\Delta}+\frac{\mu_sn_s}{2\pi\Delta^2}
    S(|k_0|a\Delta)+ 
    \left.\dot{\tilde{\omega}}_s\right|_{\mbox{\scriptsize obl}}.
\end{equation}

When the disk and satellite are in steady--state, 
the satellite and its spiral wave 
pattern both precess in concert, so $\dot{\tilde{\omega}}_s=\omega(|k|)$,
which becomes
\begin{equation}
  \label{DR0}
  \pi D'(z)|k|a\Delta=\varepsilon+
    \frac{\mu_c}{\mu_d}\left(1+\frac{x}{\Delta}\right)
    +\frac{\mu_s}{2\mu_d\Delta}f(|k_0a\Delta|, z),
\end{equation}
where
\begin{equation}
  f(|k_0a\Delta|, z)=S(|k_0|a\Delta)-
    \frac{|k_0a\Delta|^2}{(|k_0a\Delta|+z)^2}
\end{equation}
is another function that takes values of $-1\le f\le 1$, and
\begin{equation}
  \label{mu_c}
  \mu_c\equiv\frac{21\pi}{4}
    \left(\frac{R_p\Delta}{a_s}\right)^2J_2
\end{equation}
is called the critical disk mass \citep{H07}.

Two additional assumptions will then provide a simple
expression for the wavenumber $k$.
First, assume that the right term in Eqn.\ (\ref{DR0})
is small compared to the middle term, which requires
the satellite's mass to be sufficiently small, namely, that
$\mu_s\ll2\mu_c\Delta$. Second, assume that the dimensionless
wavenumber obeys $H\ll1$, which means that $D'(z)\rightarrow1$
downstream. Note, though, that $D'(z)$ is not unity
in the wave launch zone; rather it takes
values of $1/2\le D'(z)\le1$ over the initial wavelength
(see Fig.\ 2 of \citealt{H07}). However, a sufficiently
accurate result is obtained when $D'(z)$ is replaced with
its average value over the first wavelength,
$\bar{D}\simeq0.87$ \citep{H07}.
These assumptions will be confirmed below
in Section \ref{assumptions}. With these assumptions in hand, 
Eqn.\ (\ref{DR0}) is then yields the wavenumber
\begin{equation}
  \label{k}
  |k(x)|\simeq\frac{1}{\pi\bar{D}a\Delta}
    \left[\varepsilon+\frac{\mu_c}{\mu_d}
    \left(1+\frac{x}{\Delta}\right)\right],
\end{equation}
with
\begin{equation}
  \label{k0}
  |k_0|=\frac{\varepsilon+\mu_c/\mu_d}{\pi\bar{D}a\Delta}
\end{equation}
being the initial wavenumber evaluated at $x=0$.

The first wavelength $\lambda_0$ is obtained from
$\int_0^{\lambda_0}|k|da=2\pi$.
When $\mu_c\ll\mu_d$, that integral yields
\begin{equation}
  \label{lambda0}
  \lambda_0\simeq2\pi^2\bar{D}a\Delta/\varepsilon.
\end{equation}
However, when $\mu_c\gtrsim\mu_d$, (which occurs when the central planet
is too oblate or the disk mass is too small), the wavenumber $k$ 
varies with distance $x$, which violates a key assumption
in the derivation of the wave amplitude (Section \ref{ring_disk evol}).
The numerical experiment described in Section \ref{Pan} will also show 
that the wave amplitude (Eqn.\ \ref{e}, derived below)
is smaller than expected in this case. Nonetheless, that experiment
also shows that the formula for the expected wavenumber, Eqn.\ (\ref{k}),
is quite reliable when $\mu_c\gtrsim\mu_d$.

\subsubsection{checking the assumptions}
\label{assumptions}

These results will be applied to the Saturnian satellites
Pan and Daphnis that orbit in narrow gaps in Saturn's main A ring.
Pan has a mass $\mu_s=8.7\times10^{-12}$ Saturn masses
\citep{PTW07} and a semimajor axis $a_s=133,584$ km
\citep{JSP07}.
Pan orbits near the center of the Encke gap whose half-width is 
$\Delta a=162.5$ km \citep{Petal05}, so its fractional half-width is
$\Delta=\Delta a/a_s=0.0012$. Saturn's radius is  $R_p=60,330$ km
and its second zonal harmonic is $J_2=0.0163$,
so the critical disk mass is $\mu_c=7.9\times10^{-8}$.
Consequently, the quantity $2\mu_c\Delta=1.9\times10^{-10}$,
which does indeed satisfy the requirement that 
$\mu_s\ll2\mu_c\Delta$.

This same assumption is also satisfied by
Daphnis, whose mass is $\mu_s=1.5\times10^{-13}$ \citep{PTW07},
but by a smaller margin. Daphnis orbits in the Keeler gap, 
which itself is about $2900$ km
beyond the Encke gap.
The Keeler gap is also about 8 times narrower than the Encke gap
\citep{Tetal05}, so $\Delta\simeq1.5\times10^{-4}$,
which reduces the critical disk mass to
$\mu_c\simeq1.2\times10^{-9}$. Consequently,
$\mu_s<2\mu_c\Delta$ is still satisfied, but only by a factor of $\sim2.5$.

The other requirement is that $H=\sqrt{2}\mathfrak{h}|k|a<<1$.
The ring's vertical thickness is about $h\simeq30$ m near
the Encke gap \citep{Tetal07}, so the ring's fractional thickness is
$\mathfrak{h}=h/a_s\simeq2\times10^{-7}$; a comparable thickness
can also be inferred from the A-ring viscosity that is
reported in \cite{PTW07}. The ring's surface
density here is $\sigma\simeq46$ gm/cm$^2$ \citep{Tetal07},
so the normalized disk mass is
$\mu_d=\pi\sigma a_s^2/M\simeq4.5\times10^{-8}$,
which means that Pan's initial wavenumber is
$|k_0|a\simeq1100$ by Eqn.\ (\ref{k0}), while Daphnis has
$|k_0|a\simeq5000$. Consequently,
$H=\sqrt{2}\mathfrak{h}|k|a\sim{\cal O}
(10^{\mbox{\scriptsize -3 to -4}})<<1$
is very well satisfied. And since $H<1$, this means that these satellites
will launch long trailing density waves that propagate radially
outwards from the gap edge, provided these satellites have nonzero
eccentricities.

\subsubsection{waves in an interior disk}
\label{interior}

A satellite that orbits in the center of a gap in a broad planetary
ring might also excite a disturbance in the disk material that orbits interior
to the satellite. The dispersion relation for any density
waves launched at the inner gap edge is
\begin{equation}
  \label{k_in}
  |k(x)|\simeq\frac{1}{\pi\bar{D}a|\Delta|}
    \left[\varepsilon-\frac{\mu_c}{\mu_d}
    \left(1+\left|\frac{x}{\Delta}\right|\right)\right],
\end{equation}
which may be derived via the method that is described in footnote 1 of \cite{H07}.
This dispersion relation is identical to Eqn.\ (\ref{k}) except 
for the sign on the term that is proportional to the
critical mass. Since the right hand side of Eqn.\ (\ref{k_in}) 
must be positive wherever the wave propagates, 
this equation tells us that these density waves are only allowed 
inwards a distance $|x|<x_{\mbox{\scriptsize in}}$, where
\begin{equation}
  \label{x_in}
  x_{\mbox{\scriptsize in}}\equiv
    \left(\frac{\varepsilon \mu_d}{\mu_c}-1\right)|\Delta|
\end{equation}
is the distance of the waves' maximum inwards excursion.
For Pan, this distance evaluates to 
$x_{\mbox{\scriptsize in}}\simeq0.1\Delta$, 
which corresponds to a physical distance of 
$x_{\mbox{\scriptsize in}}a_s\simeq15$ km, which is only a tiny
fraction of the wave's initial wavelength that is calculated below
in Section \ref{wave-amplitude2}. In short, the A ring
material that orbits interior to Pan is unable to
sustain this type of density wave.

Note, however, that $x_{\mbox{\scriptsize in}}$ is a bit
larger for the ring material that orbits interior
to Daphnis and the Keeler gap. However, Daphnis'
small mass \citep{PTW07} and low eccentricity \citep{JSP07}
results in waves of such low amplitude
that this issue is moot.

\subsection{wave amplitude, continued}
\label{wave-amplitude2}

Plugging the initial wavenumber, Eqn.\ (\ref{k0}),
into Eqn.\ (\ref{edownstream}) then yields
an expression for the amplitude of the density wave
that is now written only in terms of the 
system's known physical parameters:
\begin{equation}
  \label{e}
  \frac{e}{e_s}\simeq
    \frac{\mu_s(\varepsilon+\mu_c/\mu_d)}{2\pi\bar{D}\mu_d \Delta}.
\end{equation}
However, keep in mind that this derivation requires that the wavelength
vary little over the first wavelength, which in turn requires
$\mu_c\ll\mu_d$. The simulations described in Section \ref{Pan}
will show that Eqn.\ (\ref{e}) overestimates the wave amplitude
when $\mu_c\gtrsim\mu_d$.

Pan's eccentricity is $e_s=1.4\times10^{-5}$
\citep{JSP07}, so Eqn.\ (\ref{e}) predicts 
eccentricities of $e\sim1.5\times10^{-6}$
due to the density wave that Pan excites at the outer edge of the
Encke gap, since $\varepsilon=2$,
$\mu_c=7.9\times10^{-8}$, and $\mu_d\simeq4.5\times10^{-8}$
(see Section \ref{assumptions}).
However, this is actually an overestimate,
since $\mu_c/\mu_d\simeq1.8$ is not small, which is a key assumption
in the derivation of Eqn.\ (\ref{e}). In fact, the simulation described
in Section \ref{Pan} will show that Eqn.\ (\ref{e}) overestimates
the amplitude of Pan's waves by a factor of $\gamma\simeq4$,
so the epicylic amplitude due to this wave
at the outer Encke gap is only $\Delta r=ea_s/\gamma\sim50$m,
which is likely too small to be seen by a spacecraft
such as Cassini. And since this wave's
initial wavenumber is $|k_0|a\simeq1100$ (Section \ref{assumptions}),
its initial wavelength would be $\lambda_0=2\pi/|k_0|\simeq760$ km,
which is also far longer than the wavelengths of any of the more
familiar density waves that satellites routinely launch at
their many mean-motion resonances in the rings.

Recall that the wavenumber $|k|$ increases with distance
$x$ (see Eqn.\ \ref{k}), so the wavelength $\lambda$ shrinks as the wave
propagates outwards. Note that the A ring's outer edge is 
$\Delta a\simeq3200$ km away from the Encke gap,
which corresponds to a fractional distance of 
$x=\Delta a/a_s\simeq0.024$.
Inserting this into Eqn.\ (\ref{k}) then yields a wavenumber
of $|k|a\simeq1.2\times10^4$, which corresponds to
a wavelength of $\lambda_0=2\pi/|k_0|\simeq70$ km when the
wave hits the outer edge of the A ring. This presumes 
that the wave was not damped en route by the ring's viscosity,
but Section \ref{viscous} will show that the viscous damping
of these long outbound waves is modest.
The reflection of these undamped waves near the outer edge of the A ring
is also discussed in Section \ref{reflect}.

The fractional variation in the disk's surface density
due to the density wave is \citep{BGT85, H03}
\begin{equation}
  \label{ds0}
  \frac{\Delta\sigma}{\sigma_0}\simeq
    \frac{\partial(ea)}{\partial a}\cos(\theta-\tilde{\omega})
    +ea\frac{\partial\tilde{\omega}}{\partial a}\sin(\theta-\tilde{\omega}),
\end{equation}
where $\theta$ is the longitude in the disk and
$\sigma_0$ is the surface density of the undisturbed disk.
The first term is negligible since the downstream eccentricity $e$
is constant, so the magnitude of the fractional
variation in surface density variations due to the wave is
dominated by the second term, which is
\begin{equation}
  \label{ds}
  \left|\frac{\Delta\sigma}{\sigma_0}\right|\simeq ea|k|
\end{equation}
since $k=-\partial\tilde{\omega}/\partial a$
(see Eqn.\ \ref{wavenumber_exact}). Since the wave launched by
Pan has an amplitude of $e\sim4\times10^{-7}$
and an initial wavenumber of $|k_0|a\simeq1100$,
the surface density variations due to this wave are quite small,
only $\Delta\sigma/\sigma_0\simeq ea|k|\sim4\times10^{-4}$, which 
again is too small for detection. However, it might be easier
to see this wave further downstream, due to the increase in $|k|$
with distance $x$. For instance, $|k|$ will have increased by a
factor of 10 when the wave reaches the outer part of the A ring,
so $\Delta\sigma/\sigma\sim4\times10^{-3}$ there, but this again is
probably still be too small for detection. 
Thus the observational consequences of these
density waves are seemingly slight. Nonetheless, these waves
are not totally inconsequential,
since Section \ref{e-damping} will show that the excitation of these
waves can also result in a vigorous damping of the satellite's
eccentricity.

\subsubsection{viscous damping of spiral density waves}
\label{viscous}

The variations in the ring's surface density due to the density
wave are dominated by the second term in Eqn.\ (\ref{ds0}), so
$\Delta\sigma\simeq-\sigma_0eka\sin(\theta-\tilde{\omega})$,
where the ring's longitude of periapse is
$\tilde{\omega}(a,t)=\omega t-\int^ak(r)dr$, with the
$\omega t$ term accounting for the spiral pattern's rotation
with time due to the system's precession. Inserting this into the above
shows that $\Delta\sigma$ has the form
$\Delta\sigma=\Re\left(Se^{i\phi}\right)$,
which has an amplitude $|S(a)|=\sigma_0e|k|a$ and a phase
\begin{equation}
  \label{phase}
  \phi(a,\theta,t)=m\theta-\omega t+\int^ak(r)dr,
\end{equation}
noting that $\theta$ in the above was multiplied by $m=1$
so that this work is readily compared to other studies of density
waves developed for spiral patterns having $m\ge1$ arms.

When there is no dissipation in the system, the wavenumber $k$ 
in Eqn.\ (\ref{phase}) is real (e.g., Eqn.\ \ref{k}). However,
if there is any dissipation in the disk, then the wavenumber $k$ 
acquires an imaginary component, {\it i.e.}, $k\rightarrow k_R+ik_I$,
which causes exponential damping
of the wave's amplitude. If that dissipation is due to the disk's 
kinematic viscosity $\nu$, then the imaginary part of the wavenumber 
is\footnote{Comparison of Eqn.\ (\ref{phase}) to the phase
convention adopted in \cite{S84} shows that the signs of $m$
and $\omega$ are reversed. This sign reversal is accounted for
in  Eqn.\ (\ref{kI}).} \citep{S84}
\begin{equation}
  \label{kI}
  k_I=\frac{\nu k_R^3}{mn-\omega}+
    \frac{7s_k\nu k_R^2(mn-\omega)}{6\pi G\sigma_0}
    \simeq\frac{\nu k_R^3}{n}\left(1+\frac{7}{6\mu_d|k_R|a}\right)
\end{equation}
where $k_R$ is the real part of the wavenumber (Eqn.\ \ref{k}),
$s_k=\mbox{sgn}(k_R)=+1$, $m=1$, and noting that
$\pi G\sigma_0=\mu_dan^2$ and that
the spiral pattern rotates slowly, {\it i.e.}, $\omega\ll n$. 
Since the long waves launched by Pan
have wavenumbers of $|k|a\sim10^{\mbox{\scriptsize 3 to 4}}$
(see Section \ref{wave-amplitude2}), and that
the A ring's normalized disk mass is $\mu_d\simeq4.5\times10^{-8}$
(Section \ref{assumptions}), 
it is clear that the second term in Eqn.\ (\ref{kI}) 
dominates over the first, so
\begin{equation}
  \label{kI2}
  k_Ia\simeq\frac{7}{6\mu_d}\left(\frac{\nu}{a^2n}\right)|k_Ra|^2
\end{equation}
for outward-propagating long waves. Consequently, the ring's surface
density varies as
\begin{equation}
  \label{delta_sigma2}
  \Delta\sigma=\Re\left(Se^{i\phi}\right)
    \propto\exp\left(-\int_0^{x}k_I(r)dr\right),
\end{equation}
which reveals how ring viscosity reduces the amplitude of the wave
as it travels a fractional distance $x$.

It is anticipated that the wave's viscous damping length $\ell_\nu$
will be much larger than the gap half-width $\Delta$. In that case,
the wavenumber that appears in the above formulae is approximately
\begin{equation}
  |k_R|a\simeq\frac{\mu_c x'}{\pi\bar{D}\mu_d\Delta^2}
\end{equation}
by Eqn.\ (\ref{k}). Inserting this into Eqn.\ (\ref{kI2})
and then evaluating the integral in Eqn. (\ref{delta_sigma2}) shows that
\begin{equation}
  \label{delta_sigma3}
  \Delta\sigma\propto e^{-(x/\ell_\nu)^3}
\end{equation}
where $\ell_\nu$ is the viscous damping length in units of $a_s$:
\begin{equation}
  \label{ell_nu}
  \ell_\nu=\left[\left(\frac{18\mu_d^3}{7}\right)\left(\frac{a^2n}{\nu}\right)
    \left(\frac{\pi\bar{D}\Delta^2}{\mu_c}\right)^2\right]^{1/3}.
\end{equation}
The following evaluates this viscous damping length $\ell_\nu$
for the waves that Pan would launch at the outer edge of the Encke gap.

There are two sources of viscosity in Saturn's rings:
collisions among ring particles, 
and gravitational wakes. Viscosity due to collisions among
ring particles is $\nu_c\simeq0.46 c^2\tau/n(1+\tau^2)$
where $c=hn$ is the particle's dispersion velocity and $\tau$
is the ring optical depth \citep{GT82}. Thus the dimensionless
viscosity combination $\nu_c/a^2n$ in the above
evaluates to
\begin{equation}
  \frac{\nu_c}{a^2n}\simeq\frac{0.46\mathfrak{h}^2\tau}{1+\tau^2}
  \simeq8\times10^{-15}
\end{equation}
since the outer A ring has an optical depth of $\tau\simeq0.6$
and $\mathfrak{h}\simeq2\times10^{-7}$.
Note that the viscosity that is associated with the ring's
gravitational wakes $\nu_g$ is also comparable, since
\begin{equation}
  \frac{\nu_g}{a^2n}=
    \frac{CG^2\sigma^2}{a^2n^4}=\frac{C\mu_d^2}{\pi^2}
    \simeq7\times10^{-15}
\end{equation}
where the coefficient $C\simeq33$ for the A ring \citep{Detal01}
and $\mu_d\simeq4.5\times10^{-8}$. This all suggests that
the total viscosity in the outer A ring is around 
$\nu/a^2n\sim10^{-14}$. Inserting this into Eqn.\ (\ref{ell_nu})
then shows that the viscous damping length for Pan's waves
(with $\Delta=0.0012$ and $\mu_c=7.9\times10^{-8}$)
is $\ell_\nu\simeq0.039$, which corresponds to a physical
distance of $\ell_\nu a_s\simeq5200$ km. Note, though,
that the outer edge of the A ring only lies a fractional distance
distance of $x\simeq0.024$, which means that the amplitude of
Pan's outbound density waves is reduced only by a factor of
$e^{-(x/\ell_\nu)^3}\simeq0.8$ as it travels across the A ring. 
This calculation indicates that the damping of long waves
due to ring viscosity is only of marginal importance.

Section \ref{Pan} will also show
that this long wave will reflect 
at the outer edge of the A ring (or else at the nearby 
Keeler gap) and march back towards Pan
as a superposition of both long and short waves that are of roughly 
equal magnitudes. 
Pan's outbound long density waves will have a dimensionless wavenumber
$H_L=\sqrt{2}\mathfrak{h}|k|a\simeq3.4\times10^{-3}\ll1$ when they near
the outer edge of the A ring.  Let $H_S$
be the dimensionless wavenumber of the reflected 
short waves. Since the long and
short spiral wave patterns both precess at the same rate,
the wavenumbers $H_L$ and $H_S$ both satisfy the same
dimensionless dispersion relation, Eqn.\ (\ref{dim_disp_rel}), so
\begin{equation}
  H_L\simeq\omega^\star=H_Se^{-H_S}.
\end{equation}
This is solved numerically for the wavenumber of the reflected short
wave, $H_S\simeq7.7$, which yields a physical wavenumber of
$|k_s|a\simeq2.7\times10^7$ since $\mathfrak{h}\simeq2\times10^{-7}$.
This corresponds to a wavelength 
$\lambda_s=2\pi/|k_s|$ that is comparable to the 
disk's scale height $h=\mathfrak{h}a_s\sim30$m. 
These very short-wavelength waves
will be nonlinear ($\Delta \sigma/\sigma>1$), and they will likely
damp on a very short spatial scale.

\section{Damping the satellite's eccentricity}
\label{e-damping}

The excitation of these density waves also alters the satellite's
eccentricity $e_s$ at a rate that can be calculated from
the angular momentum flux that is transported by these waves.
The angular momentum content of a narrow annulus in the disk
is $\delta L=\delta m\sqrt{GMa(1-e^2)}$, where
$\delta m=\sigma\delta A$ is the mass of that annulus
which has an area $\delta A$ and a mass surface density $\sigma$. The
surface density  of angular momentum in that annulus is then
$\ell=\delta L/\delta A=\sigma na^2\sqrt{1-e^2}
\simeq\sigma na^2(1-e^2/2)=\ell_0+\ell_w$, where
$\ell_0=\sigma na^2$ would be the angular momentum surface density 
if the ring were circular, and $\ell_w=-\sigma e^2na^2/2$ is the
surface density of angular momentum that is associated with
the wave whose amplitude is $e$. The flux of angular momentum
that is transported by the wave is $F=\ell_w c_g$,
where $c_g=\mu_d an$ is the group velocity of long density waves
(see Section \ref{group velocity}). Consequently, the 
angular momentum luminosity, which is the rate at which
waves transport angular momentum across an annulus of radius $a$, is
${\cal L}=2\pi a F=-(e\mu_dan)^2M$. 

Note the sign on ${\cal L}$, which means that these waves transport
angular momentum inwards, from the disk to the satellite. This transport
also increases the satellite's angular momentum 
$L_s=m_s\sqrt{GMa_s(1-e_s^2)}\simeq m_s n_s a_s^2(1-e_s^2/2)$ 
at the rate $dL_s/dt\simeq-m_s n_s a_s^2 e_s\dot{e}_s=-{\cal L}$,
which then provides the rate at which wave excitation tends to damp
the satellite's eccentricity:
\begin{equation}
  \label{edamping1}
  \dot{e}_s=-\frac{e^2\mu_d^2}{e_s\mu_s}n_s.
\end{equation}
And if the wavenumber varies little over that first wavelength,
which requires $\mu_c\ll\mu_d$, then the wave amplitude
$e$ is given by Eqn.\ (\ref{e}),
which yields the satellite's eccentricity damping rate
\begin{equation}
  \label{edamping2}
  \frac{\dot{e}_s}{e_s}=
    -\frac{\mu_s(\varepsilon+\mu_c/\mu_d)^2}{(2\pi\bar{D}\Delta)^2}n_s
\end{equation}
and the eccentricity decay timescale
\begin{equation}
  \label{tau_e}
  \tau_e=\left|\frac{e_s}{\dot{e}_s}\right|=
    \frac{2\pi(\bar{D}\Delta)^2}{\mu_s(\varepsilon+\mu_c/\mu_d)^2}
    P_{\mbox{\scriptsize orb}},
\end{equation}
where $P_{\mbox{\scriptsize orb}}=2\pi/n_s$ 
is the satellite's orbit period.

If, however, $\mu_c\gtrsim\mu_d$, then Eqn.\ (\ref{e}) will
overestimate the wave amplitude $e$, and Eqn.\ (\ref{edamping2}) 
will not be valid.  Nonetheless, Eqn.\ (\ref{edamping1})
can still be used to determine the satellite's $e$-damping rate, 
but the wave amplitude $e$ must be determined by other means,
such as by using the rings model that is described in Section \ref{sims}.

\subsection{comparison to Lindblad and corotation resonances}
\label{resonances}

The satellite is also perturbing the planetary ring
at its many Lindblad and corotation resonances in the disk,
and this interaction also causes the satellite's eccentricity
to evolve at the rate
\begin{equation}
  \left.\dot{e}_s\right|_{LC}=\left.\dot{e}_s\right|_{L}+
    \left.\dot{e}_s\right|_{C}
\end{equation}
where the first term represents the eccentricity excitation that is
due to the Lindblad resonances, and the second term
is the eccentricity damping that is due to 
corotation resonances. The rate at which the satellite's eccentricity $e_s$
evolves due to its Lindblad resonances  
in a narrow annulus of mass $\delta m'$ that lies a 
fractional distance $x'$ away is
\begin{equation}
  \label{delta_es_LR}
  \left.\delta\dot{e}_s\right|_L=
    \frac{f_L\mu_se_s}{2|x'|^5}\frac{\delta m'}{M}n_s
\end{equation}
(from \citealt{GT81}), 
where the factor $f_L\simeq3.045$. The satellite's total eccentricity
variation due to all of its Lindblad resonances throughout the
disk is obtained by
setting $\delta m'=2\pi\sigma' a'da'=2\mu_d' Mdx'$ and
integrating Eqn.\ (\ref{delta_es_LR}) across the disk, which yields
\begin{equation}
  \label{es_LR}
  \left.\dot{e}_s\right|_L=2\int_\Delta^\infty
    \left.\delta\dot{e}_s\right|_L=
    \frac{f_L\mu_s\mu_de_s}{2\Delta^4}n_s,
\end{equation}
where the factor of two on the middle term accounts for the ring material
orbiting interior and exterior to the satellite. The $e$-excitation
timescale due to Lindblad resonances is then
\begin{equation}
  \label{tau_L}
  \tau_L=\frac{e_s}{\left.\dot{e}_s\right|_L}=
    \frac{\Delta^4}{\pi f_L\mu_s\mu_d}
    P_{\mbox{\scriptsize orb}}.
\end{equation}

The satellite also has many corotation resonances 
that lie in the planetary ring,
and their effect is to damp the satellite's eccentricity at a
rate $\left.\dot{e}_s\right|_C$ that
has the same form as Eqns.\ (\ref{delta_es_LR}--\ref{es_LR}) but with a 
lead coefficient of $f_C=-3.193$
\citep{GT81}. However this eccentricity
damping only occurs if the particles' motions at the 
corotation resonances are not saturated.
The total rate at which the satellite's eccentricity varies
due to its Lindblad and corotation resonances is
\begin{equation}
  \label{es_LR+CR}
  \left.\dot{e}_s\right|_{LC}=\left.\dot{e}_s\right|_{L}+
    \left.\dot{e}_s\right|_{C}
    =\frac{f_{LC}\mu_s\mu_de_s}{2\Delta^4}n_s
\end{equation}
where the factor $f_{LC}=f_L+f_C=-0.148$, 
which indicates that the net effect of these resonances
is to damp the satellite's eccentricity, provided the
corotation resonances are not saturated.
Note, though, that $|f_{LC}/f_L|\simeq5\%$,
which means that eccentricity damping wins by only a small margin.
Consequently, the total $e$-damping
timescale due to Lindblad and corotation resonances is then
\begin{equation}
  \label{tau_LC}
  \tau_{LC}=\left|\frac{e_s}{\left.\dot{e}_s\right|_{LC}}\right|=
    \frac{\Delta^4}{\pi |f_{LC}|\mu_s\mu_d}
    P_{\mbox{\scriptsize orb}}.
\end{equation}

If, however, the corotation resonances are saturated,
then the Lindblad resonances will pump up the satellite's
eccentricity, but at a rate that is half of Eqn.\ (\ref{es_LR})
(see \citealt{GT81}).
Nonetheless, the satellite's secular perturbations of the ring
will also endeavor to damp the satellite's
eccentricity at the rate given by
Eqn.\ (\ref{edamping2}). Comparing these two rates will show
that eccentricity damping is still assured when
the satellite's gap is sufficiently wide:
\begin{equation}
  \label{gap_width}
  \Delta^2>f_L(\pi\bar{D})^2\mu_d/2,
\end{equation}
provided $\mu_c\ll\mu_d$.

Daphnis satisfies the $\mu_c\ll\mu_d$ condition, and since
$\mu_d\simeq4.5\times10^{-8}$, $\bar{D}=0.87$,
and $\varepsilon=2$, the above gap-width requirement
evaluates to $\Delta\gtrsim7\times10^{-4}$.
However the Keeler gap's half width is only
$\Delta\simeq1.5\times10^{-4}$, so this requirement is not
satisfied. Consequently,
if Daphnis' corotation resonances were in fact saturated, then
this secular $e$-damping could not counterbalance
the effects of the Lindblad resonances, which would pump up 
Daphnis' eccentricity until that satellite crashed into the A ring.
That Daphnis has a very low eccentricity
suggests that the motions of the 
ring particles at its corotation resonances are not saturated.

Note that Pan does not satisfy $\mu_c\ll\mu_d$,
so Eqn.\ (\ref{gap_width}) does not apply to this satellite. Instead,
the rings model of Section \ref{Pan} is used to assess
that satellite's $e$-damping rate.

\section{Simulations of density waves launched at a gap edge}
\label{sims}

The rings model of \citet{H03} is used to test the preceding results.
This model treats the disk as a set of 
N discrete gravitating annuli having semimajor axes $a_j$, eccentricities 
$e_j$,  longitudes of periapse  $\tilde{\omega}_j$, and
half--thicknesses $h_j$. 
The model only considers the system's secular 
gravitational perturbations, so it also solves the same equations of motion,
Eqns.\ (\ref{edot}), but it does so without making any of 
the approximations and assumptions
invoked in Section \ref{EOM}. Consequently, the 
model provides another check on the analytic
results obtained above. 

\subsection{outbound waves}
\label{outbound waves}

The rings model is used to simulate the spiral density waves that are
launched by an eccentric satellite that orbits just interior to a disk.
Figure \ref{ewave} shows the amplitude of this wave 
as it advances across a disk, with the system's parameters being
detailed in the figure caption. Note, though, that those 
parameters do not correspond  to any real ring--satellite system. Rather, 
these parameters were chosen to illustrate the results of
Section \ref{EOM} in the limit in which those results
were obtained, namely, that the factor $\mu_c/\mu_d$
appearing in the wavenumber Eqn.\ (\ref{k}) is modest,
with $\mu_c/\mu_d=0.2$. This causes
the wavenumber to vary only a small amount across the first wavelength,
which is a key assumption of Section 
\ref{amplitude}'s derivation of the wave amplitude.
Nonetheless, the simulation reported in Fig. \ref{ewave} 
does correspond loosely to a small $\sim10$ km satellite orbiting 
just interior  to a ring whose surface density is similar to 
Saturn's main A ring.

The amplitude of the simulated wave compares favorably with 
the wave's expected amplitude, Eqn.\ (\ref{e}), which is indicated by
the dashed line in Fig.\ \ref{ewave}.
The wave propagation time also provides another check on these
calculations. Note that the time for these waves to propagate a 
fractional radial distance $x=\Delta r/a$ is
\begin{equation}
  t_{prop}=\frac{\Delta r}{c_g}=\frac{xP_{orb}}{2\pi\mu_d},
\end{equation}
where $c_g$ is the waves' group velocity, Eqn.\ (\ref{cg}).
The simulated disk has a normalized mass of $\mu_d=5\times10^{-8}$
and a fractional width $x=0.015$, so the anticipated propagation time
is $t_{prop}=48\times10^3$ orbits, which compares well with with the
simulated wave's propagation time (Fig. \ref{ewave}).

Section \ref{group velocity} predicts that the satellite will launch 
a long trailing density wave that has $s_k=\mbox{sgn}(k)=+1$.
So by Eqn.\ (\ref{wavenumber_exact}), this means that the disk's
longitude of periapse $\tilde{\omega}(a)$ 
should steadily decrease at greater distances $a$ in the disk.
This is confirmed in Fig.\ \ref{wk}, which 
shows the waves' longitudes relative to the satellite's,
$\tilde{\omega}(a)-\tilde{\omega}_s$. 
Also note that the longitude of periapse at the disk's inner edge is 
$90^\circ$ behind the satellite's longitude, as
expected. We find that once the density wave is established in the
disk, its longitudes $\tilde{\omega}(a)$
precess at the same rate as the satellite's,
and that the disk's eccentricities $e(a)$ are also constant, 
which confirms the  steady--state assumption employed in 
Eqns.\ (\ref{ss}).
The disk's surface density variations due to the wave is Eqn.\ (\ref{ds0}),
which is dominated by the second term. Inspection of that term
shows that the disk's maximum surface density occurs at longitudes
that trail those shown in Fig.\ \ref{wk} by $90^\circ$, while the
disk's minimum surface density Fig.\ \ref{wk} leads by $90^\circ$.

Figure \ref{wk} also plots the dimensionless wavenumber $|k|a\Delta$ across
the disk at time $t=60\times10^{3}$, when the density wave just starts 
to reflect at the disk's outer edge. Also plotted
is the expected wavenumber, Eqn.\ (\ref{k}), which compares well.
The simulation's variations in $k$ at the disk's far edge are due
to the wave reflecting there, while
the variations near the disk's inner edge
are due to the very short wavelength 
variations in $\tilde{\omega}(a)$ that are just barely seen in the upper 
figure. Those wiggles are the short waves that are described
in Section \ref{group velocity}. Evidently,
the satellite also launches short-wavelength leading waves
at the disk's inner edge, which are the cause for the 
high-frequency wiggles also seen in the 
$t=60$ curve shown in Fig.\ \ref{ewave}. 
Figure \ref{wk} also shows that the short waves propagate much more
slowly than the long waves, as expected.
Note, though, that nonlinear effects that are not modeled here would
damp these nonlinear short waves on a very short spatial scale
(e.g., Section \ref{viscous}).

The rate at which the disk damps the simulated 
satellite's eccentricity $e_s$ is shown
in Fig.\ \ref{edot_fig}, which compares well with the 
expected rate, Eqn.\ (\ref{edamping2}). 

The numerical quality of this simulation is assessed by
monitoring the systems' total angular momentum deficit, which is
$L_e=\onehalf\sum_j m_j n_j a_j^2 e_j^2$,
where the sum runs over all rings and satellites in the
system. Since this quantity is conserved
by Eqns.\ (\ref{EOM}) \citep{H03}, it provides
a useful check on the simulation's numerical precision. 
The single--precision calculation shown in Fig.\ \ref{ewave}
conserves $L_e$ with a fractional error of
$|\Delta L_e/L_e|<4\times10^{-5}$.

\subsection{reflection at an outer boundary}
\label{reflect}

Section \ref{viscous} showed that
ring viscosity is only marginally effective at damping the long waves
that an eccentric satellite would launch at the gap's outer edge. 
Consequently, these waves will propagate outwards until 
they hit a barrier, such as another gap in the ring, 
or the ring's outer edge. The subsequent
fate of such a wavetrain is illustrated in 
Fig.\ \ref{ew_fig}, which shows the state of the simulation
described in Section \ref{outbound waves} but at a later time,
$t=2.1\times10^5$ orbits.
Here, the outbound long wave has already
reflected at simulated ring's outer edge and 
propagated back towards the satellite, but 
as the superposition of a long leading wave and a short trailing wave.
Consequently, the long-wavelength undulations seen in Fig.\ \ref{ew_fig}'s
$e(a)$ curve represents the superposition of outbound and inbound
long waves. Note, though, that the reflected waves' angular momentum
content is now shared between a long and a short wave,
so the amplitude of the inbound long wave is smaller than the outbound
long wave. Consequently, the broad 
$e(a)$ undulations seen in Fig.\ \ref{ew_fig}
are not due to a standing wave in the disk, but instead 
represent the superposition
of two traveling waves having different amplitudes.

The high-frequency variations seen in Fig.\ \ref{ew_fig}'s $e(a)$ curve
are due to short waves that
have a wavelength that is comparable to the disk scale height $h$.
Note that the right-hand side of Fig.\ \ref{ew_fig} 
shows that the reflected short wave is 
still confined to the vicinity of the ring's outer edge by time 
$t=2.1\times10^5$ orbits, which indicates that these short
waves travel slower than the long waves, as expected. 
The rapid variations in $e(a)$ seen at the left side of
Fig.\ \ref{ew_fig} shows that the satellite is also exciting
short waves at the gap's inner edge. These short waves are propagating much
more slowly (see Section \ref{group velocity}), by a factor of $He^{-H}\sim0.01$,
so these very slow-moving
waves have had little opportunity to travel very far by time 
$t=2.1\times10^5$ orbits. Longer-term simulations also show
that the amplitude of any short waves is always
comparable to the amplitude of the long waves.
Equation (\ref{ds}) also indicates that these short waves will
be very nonlinear ($|\Delta\sigma/\sigma|>1$), so the reflected short wave
is expected to damp over a very short spatial scale that is probably
comparable to the ring's scale height $h$,
in the vicinity of the ring's outer edge. 

However, the reflected long wave will still return to the disk's inner edge,
where it can interact with the satellite and/or reflect again.
The satellite can interact with this returning wave by absorbing 
some of the wave's angular momentum content, which would also
excite the satellite's eccentricity and seemingly
stall any further $e$-damping. However that $e$-pumping
would then be counterbalanced by enhanced eccentricity damping
due to the excitation of even higher-amplitude
density waves. Consequently, the damping of the satellite's
eccentricity by this phenomenon is still assured, 
despite the fact that long waves propagate with only a modest
amount of viscous damping.
This is due to the fact that all 
long waves eventually reflect somewhere in the
disk and spawn short waves that are easily damped.

\subsection{waves launched by Pan}
\label{Pan}

The rings code is also used to simulate the density waves that
an eccentric Pan can launch at the outer edge of the Encke gap.
As Section \ref{wave-amplitude2} notes, the ratio of the critical
mass $\mu_c$ to the disk mass $\mu_d$
is $\mu_c/\mu_d\simeq1.8$, which indicates
that the wavenumber changes substantially over the first
wavelength (see Eqn.\ \ref{k}) which violates a key assumption in
the derivation of the wave amplitude. In fact, a comparison
of Eqn.\ (\ref{e}) 
to simulations of Pan's waves shows that that equation
overestimates the wave amplitude by a factor of $\gamma\simeq4$,
which means that Eqn.\ (\ref{edamping2}), which provides the satellite's
eccentricity-damping rate, would also be in error by a factor
of $\gamma^2$. Nonetheless, a comparison of the simulated wave's
wavenumber $|k|$ to Eqn.\ (\ref{k}) shows that equation to be in 
excellent agreement with the model results. 
Eqn.\ (\ref{edamping1}) is also shown to be a reliable indicator of the
satellite's eccentricity damping rate, even when $\mu_c>\mu_d$.

These same simulations also show that density waves launched by
Pan are not able to propagate across the Keeler gap,
which lies about 2900km downstream of the Encke gap. 
The width of that gap is
approximately 40km, which is about half wavelength of Pan's waves
in this region. Despite having a wavelength that is larger than the
gap's full width, Pan's waves are unable to propagate
across the Keeler gap. Instead, those waves
reflect at the gap's inner edge, which, in the simulation,
propagate back to Pan as a superposition of long and short waves.
However in a real disk, those nonlinear
short waves would quickly damp very
near the Keeler gap's inner edge.

This wave-action will also damp Pan's eccentricity over a timescale
equal to Eqn.\ (\ref{tau_e}) multiplied by $\gamma^2=16$,
which evaluates to $\tau_e\simeq1400$ years. 
Note that this $e$-damping is  competitive with,
but not quite faster than, the eccentricity
excitation that is due to Pan's Lindblad resonances
in the ring, which pump up the satellite's $e$
over a slightly faster timescale of $\tau_L\simeq900$ years
(see Eqn.\ \ref{tau_L}).
But if the  particles' motions at Pan's corotation resonances are
saturated, then $e$-damping by the corotation torque is shut off, while 
the e-excitation due to the Lindblad torque is halved \citep{GT81},
so $\tau_L\rightarrow1800$ years. So when  Pan's corotation resonances
are saturated, Pan's $e$-damping due to its secular interaction with
the ring will exceed the $e$-excitation that is due to its Lindblad
resonances in the ring, but only by a small margin. 
But if particle motions at Pan's corotation resonances
are in fact unsaturated, then the near cancellation of the Lindblad
and corotation torques results in a secular $e$-damping
timescale that is about 13 times shorter
than the total resonant $e$-damping
timescale (Eqn.\ \ref{tau_LC}). 
In this case, $e$-damping by the
secular interaction is the dominant process
that stabilizes Pan's eccentricity.

Since this eccentricity damping due to wave excitation
is so vigorous, one might wonder why Pan would even
have a nonzero eccentricity. However,
as \cite{Setal06} point out, the satellite 
Prometheus has a 16:15 resonance that does disturb Pan, 
which may be responsible for sustaining Pan's eccentricity
and the density waves that that satellite would launch at the outer
edge of the Encke gap.

\section{Summary and Conclusions}
\label{summary}

The Lagrange planetary equations are used to study the secular 
evolution of a small planetary satellite as it orbits within
a narrow gap in a broad, self-gravitating planetary ring.
These equations show that an eccentric satellite's
secular perturbations of the nearby gap edge tend to excite
very long-wavelength spiral density waves that propagate
out to greater distances in the ring. 
These results are applied to the two small
Saturnian satellite's Pan and Daphnis, which inhabit narrow
gaps in the main A ring. It is shown that these satellites
can launch very low-amplitude ($\Delta\sigma/\sigma<0.4\%$)
long waves whose wavelengths would be of order $\lambda\sim100$'s of km.
The wavelength of these waves also shrinks with distance
due to the central planet's oblateness, which causes the spiral
pattern to wind up as the waves propagate. It is also shown that
these long waves suffer only a modest amount of viscous damping
as they propagate towards the A ring's outer edge.
A dispersion relation is derived for these waves,
which shows that a gap-embedded satellite can also excite short
waves whose wavelength is comparable to the ring's scale height $h$.
However these short waves are very nonlinear 
($\Delta\sigma/\sigma\sim1$), and will damp soon after their excitation.

The Lagrange planetary equations are also used to derive
the amplitude and wavelength of the long waves, as well as the
rate at which this wave excitation tends to damp the satellite's
eccentricity. However these analytic results are only valid
when the wavenumber $k$ varies slowly across the first wavelength,
which requires the so-called critical mass 
$\mu_c$ to be sufficiently small. Equation (\ref{mu_c})
shows that that occurs when the gap is 
sufficiently narrow,  or when the central planet's
oblateness is sufficiently small.
If, however, this requirement is not satisfied, then
the ring's model of \cite{H03} can still be used to determine
these waves' properties, and  the rate at which this
wave action also damps the satellite's eccentricity.
Note that the amplitude of these waves is proportional
to the satellite's eccentricity, so this $e$-damping also 
tends to terminate subsequent wave generation.

The rings model shows that these undamped long waves will
eventually reflect at the ring's outer edge (or at
another gap in the ring), which then spawns
both long and short waves that propagates inwards. 
Since wave reflections eventually
transmute all long waves into easily-damped short waves,
this process communicates the wave's angular momentum
content to the ring itself, which also insures
that this wave phenomenon ultimately damps
the satellite's eccentricity, too.

This eccentricity damping due to wave excitation is then
compared to the $e$-evolution rates that are due to
the satellite's interaction with ring material orbiting at
its Lindblad resonances (which tends to pump up the satellite's
eccentricity) and corotation resonances (which tend to
damp the satellite's eccentricity). It is shown
that $e$-damping due to wave excitation is the dominant process
when the gap width is sufficiently wide (see Eqn.\ \ref{gap_width}).
For the case of Pan, these $e$-damping and $e$-excitation 
rates are all comparable to each other, but Daphnis' 
$e$-evolution is dominated by the Lindblad and corotation
torques. And since Daphnis' long-term orbital stability 
requires $e$-damping to dominate over $e$-excitation,
these results also imply that particle motions at 
Daphnis' corotation resonances are unsaturated, which is
necessary for the corotation torque to be operative here.

\acknowledgments

\begin{center}
  {\bf Acknowledgments}
\end{center}

This work was supported by grant NNX07--AL446 issued by
NASA's Science Mission Directorate via its Outer Planets Research 
Program. The author thanks Carolyn Porco
for her comments on this work, and Jayme Derrah for composing Figure 1.


\begin{thebibliography}{20}
\expandafter\ifx\csname natexlab\endcsname\relax\def\natexlab#1{#1}\fi

\bibitem[{{Abramowitz} \& {Stegun}(1972)}]{AS72}
{Abramowitz}, M. \& {Stegun}, I.~A. 1972, {Handbook of Mathematical Functions}
  (Handbook of Mathematical Functions, New York: Dover, 1972)

\bibitem[{{Binney} \& {Tremaine}(1987)}]{BT87}
{Binney}, J. \& {Tremaine}, S. 1987, {Galactic dynamics} (Princeton, NJ,
  Princeton University Press, 1987, 747 p.)

\bibitem[{{Borderies} {et~al.}(1985){Borderies}, {Goldreich}, \&
  {Tremaine}}]{BGT85}
{Borderies}, N., {Goldreich}, P., \& {Tremaine}, S. 1985, Icarus, 63, 406

\bibitem[{{Brouwer} \& {Clemence}(1961)}]{BC61}
{Brouwer}, D. \& {Clemence}, G.~M. 1961, {Methods of celestial mechanics} (New
  York: Academic Press, 1961)

\bibitem[{{Daisaka} {et~al.}(2001){Daisaka}, {Tanaka}, \& {Ida}}]{Detal01}
{Daisaka}, H., {Tanaka}, H., \& {Ida}, S. 2001, Icarus, 154, 296

\bibitem[{{Goldreich} \& {Sari}(2003)}]{GS03}
{Goldreich}, P. \& {Sari}, R. 2003, \apj, 585, 1024

\bibitem[{{Goldreich} \& {Tremaine}(1981)}]{GT81}
{Goldreich}, P. \& {Tremaine}, S. 1981, \apj, 243, 1062

\bibitem[{{Goldreich} \& {Tremaine}(1982)}]{GT82}
---. 1982, \araa, 20, 249

\bibitem[{{Hahn}(2003)}]{H03}
{Hahn}, J.~M. 2003, \apj, 595, 531

\bibitem[{{Hahn}(2007)}]{H07}
---. 2007, \apj, 665, 856

\bibitem[{{Jacobson} {et~al.}(2008){Jacobson}, {Spitale}, {Porco}, {Beurle},
  {Cooper}, {Evans}, \& {Murray}}]{JSP07}
{Jacobson}, R.~A., {Spitale}, J., {Porco}, C.~C., {Beurle}, K., {Cooper},
  N.~J., {Evans}, M.~W., \& {Murray}, C.~D. 2008, \aj, 135, 261

\bibitem[{{Murray} \& {Dermott}(1999)}]{MD99}
{Murray}, C.~D. \& {Dermott}, S.~F. 1999, {Solar system dynamics} (Cambridge
  University Press)

\bibitem[{{Porco} {et~al.}(2005){Porco}, {Baker}, {Barbara}, {Beurle},
  {Brahic}, {Burns}, {Charnoz}, {Cooper}, {Dawson}, {Del Genio}, {Denk},
  {Dones}, {Dyudina}, {Evans}, {Giese}, {Grazier}, {Helfenstein}, {Ingersoll},
  {Jacobson}, {Johnson}, {McEwen}, {Murray}, {Neukum}, {Owen}, {Perry},
  {Roatsch}, {Spitale}, {Squyres}, {Thomas}, {Tiscareno}, {Turtle}, {Vasavada},
  {Veverka}, {Wagner}, \& {West}}]{Petal05}
{Porco}, C.~C., {Baker}, E., {Barbara}, J., {Beurle}, K., {Brahic}, A.,
  {Burns}, J.~A., {Charnoz}, S., {Cooper}, N., {Dawson}, D.~D., {Del Genio},
  A.~D., {Denk}, T., {Dones}, L., {Dyudina}, U., {Evans}, M.~W., {Giese}, B.,
  {Grazier}, K., {Helfenstein}, P., {Ingersoll}, A.~P., {Jacobson}, R.~A.,
  {Johnson}, T.~V., {McEwen}, A., {Murray}, C.~D., {Neukum}, G., {Owen}, W.~M.,
  {Perry}, J., {Roatsch}, T., {Spitale}, J., {Squyres}, S., {Thomas}, P.,
  {Tiscareno}, M., {Turtle}, E., {Vasavada}, A.~R., {Veverka}, J., {Wagner},
  R., \& {West}, R. 2005, Science, 307, 1226

\bibitem[{{Porco} {et~al.}(2007){Porco}, {Thomas}, {Weiss}, \&
  {Richardson}}]{PTW07}
{Porco}, C.~C., {Thomas}, P.~C., {Weiss}, J.~W., \& {Richardson}, D.~C. 2007,
  Science, 318, 1602

\bibitem[{{Shu}(1984)}]{S84}
{Shu}, F.~H. 1984, in IAU Colloq. 75: Planetary Rings, ed. R.~{Greenberg} \&
  A.~{Brahic}, 513--561

\bibitem[{{Spitale} {et~al.}(2006){Spitale}, {Jacobson}, {Porco}, \&
  {Owen}}]{Setal06}
{Spitale}, J.~N., {Jacobson}, R.~A., {Porco}, C.~C., \& {Owen}, Jr., W.~M.
  2006, \aj, 132, 692

\bibitem[{{Tiscareno} {et~al.}(2007){Tiscareno}, {Burns}, {Nicholson},
  {Hedman}, \& {Porco}}]{Tetal07}
{Tiscareno}, M.~S., {Burns}, J.~A., {Nicholson}, P.~D., {Hedman}, M.~M., \&
  {Porco}, C.~C. 2007, Icarus, 189, 14

\bibitem[{{Tiscareno} {et~al.}(2005){Tiscareno}, {Hedman}, {Burns}, {Porco},
  {Weiss}, \& {Murray}}]{Tetal05}
{Tiscareno}, M.~S., {Hedman}, M.~M., {Burns}, J.~A., {Porco}, C.~C., {Weiss},
  J.~W., \& {Murray}, C.~D. 2005, AGU Fall Meeting Abstracts, B245+

\bibitem[{{Toomre}(1964)}]{T64}
{Toomre}, A. 1964, \apj, 139, 1217

\bibitem[{{Toomre}(1969)}]{T69}
---. 1969, \apj, 158, 899

\end{thebibliography}

\newpage

\begin{figure}
\epsscale{1.0}
\plotone{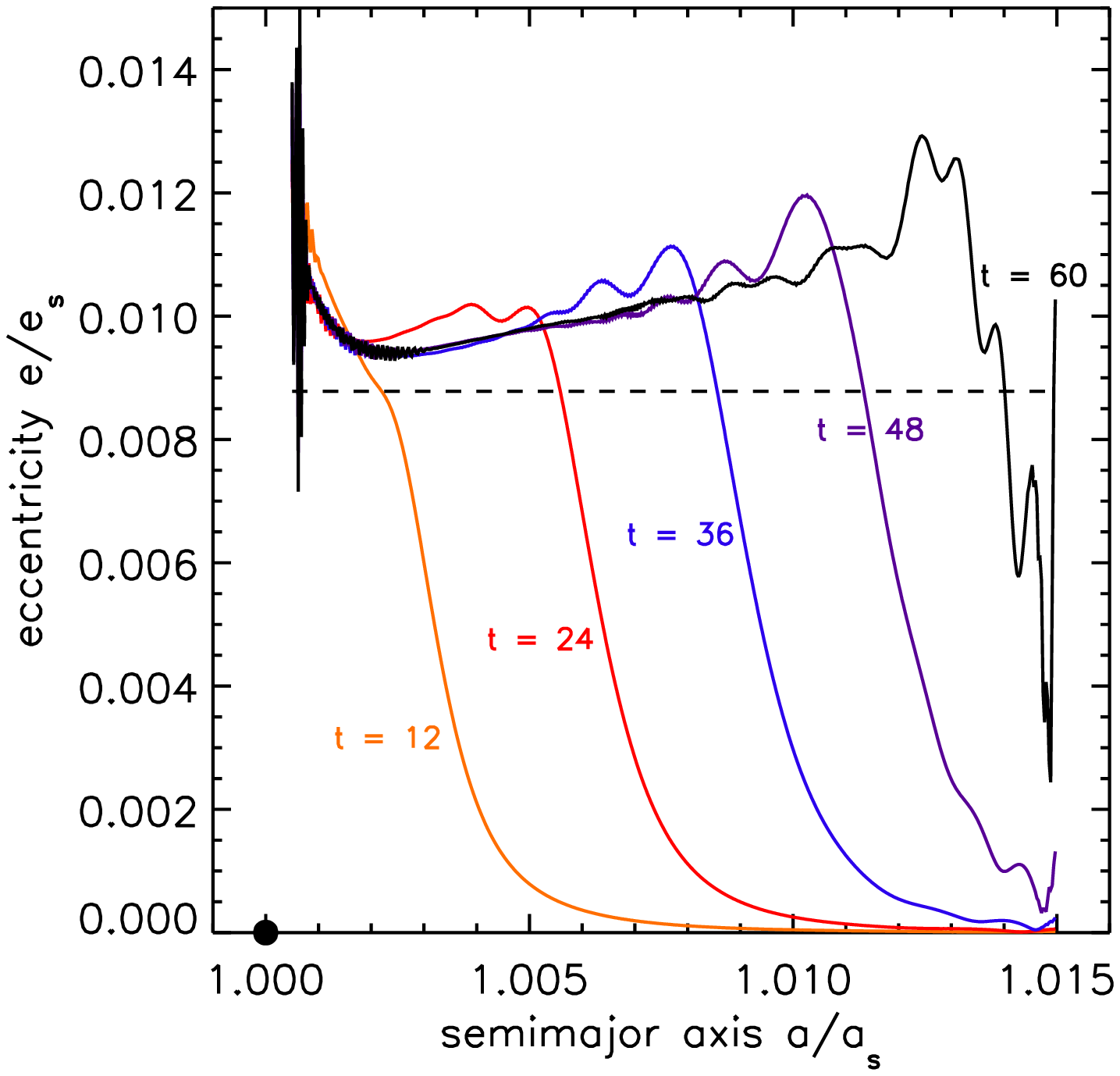}
\end{figure}

\begin{figure}
\figcaption{
  \label{ewave}
The rings model is used to simulate spiral density waves
launched by an eccentric satellite that orbits just interior to a disk.
The satellite's normalized mass is $\mu_s=10^{-12}$, 
and the disk is comprised of $N=500$ rings having semimajor axes distributed
over $1+\Delta\le a_j/a_s\le 1.015$, where $\Delta=5\times10^{-4}$ 
is the fractional distance between the 
satellite and the innermost ring. The rings' fractional masses are
$\mu_r=2.9\times10^{-12}$, so the normalized disk mass is
$\mu_d=\pi\sigma r^2/M=(\mu_r/2)(a_s/\delta)=5\times10^{-8}$,
where the rings' fractional separations are 
$\delta/a_s=0.0145/N=2.9\times10^{-5}$.
The rings' fractional half--widths $\mathfrak{h}$ is also set equal to
their separations $\delta/a_s$. The central planet's zonal 
harmonic is $J_2=0.012$ and the planet's radius
is $R_p/a_s=0.45$, so this system's critical disk mass is
$\mu_c=1.0\times10^{-8}$ and $\mu_c/\mu_d=0.2$.
The satellite's initial eccentricity is 
$e_s=10^{-5}$, with all other rings initially having
zero eccentricities.  The curves show the 
fractional amplitude of the density wave,
$e(a)/e_s$, as it advances across the disk, 
shown at selected times $t$ in units  of $10^3$ orbital periods. 
The dashed line is the expected wave amplitude,
Eqn.\ (\ref{e}), with $\varepsilon=1$. 
}
\end{figure}

\begin{figure}
\epsscale{1.0}
\plotone{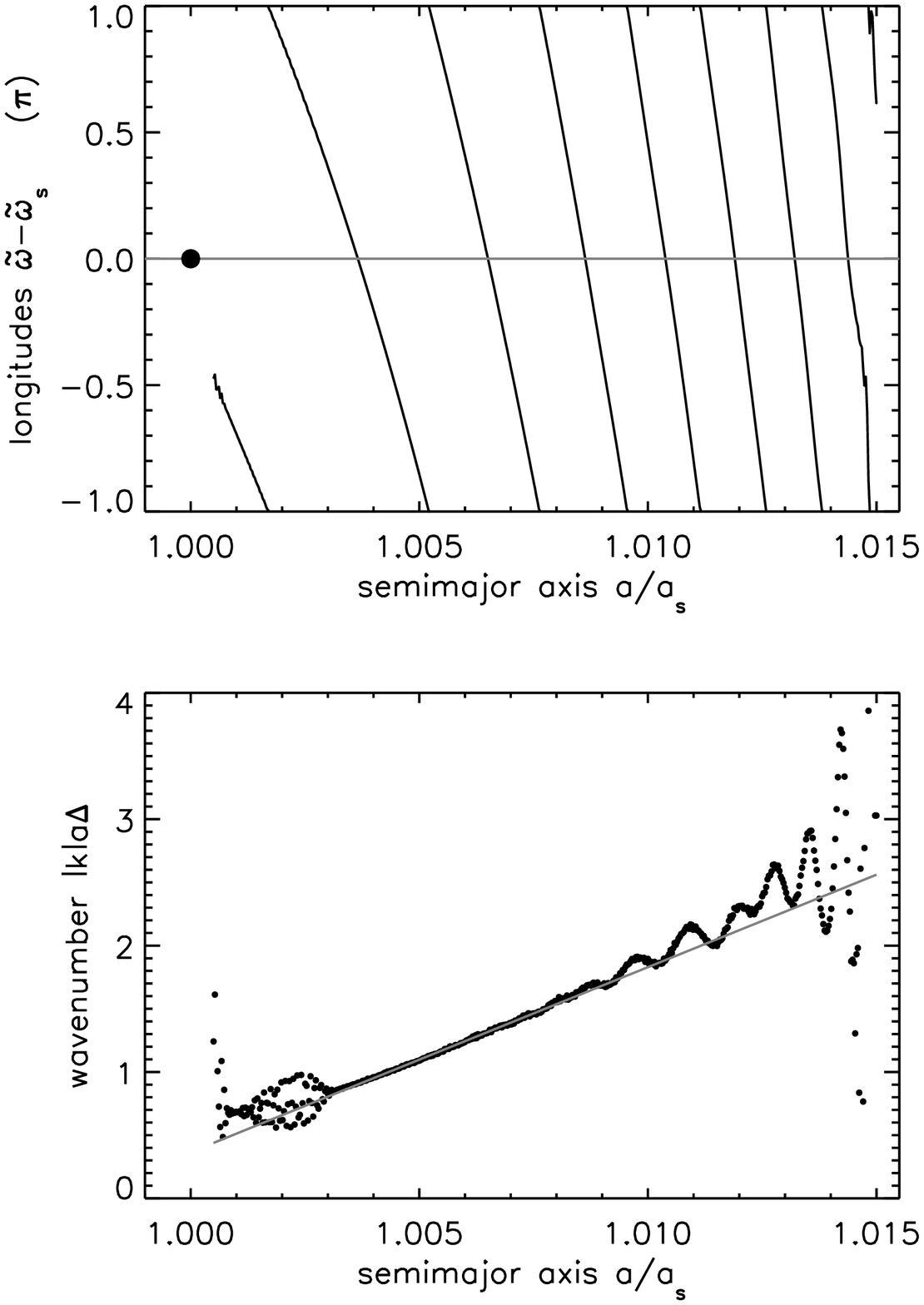}
\end{figure}

\newpage
\begin{figure}
\figcaption{
  \label{wk}
  The upper figure shows the disk's longitude of periapse
  $\tilde{\omega}(a)$ relative to the satellite's periapse 
  $\tilde{\omega}_s$, in units of $\pi$,
  for the simulation of Fig.\ \ref{ewave} at time $t=60\times10^3$ orbits,  
  when the wave has swept across the disk.
  The dots in the lower figure shows the dimensionless 
  wavenumber $|k|a\Delta$ 
  at this moment, where wavenumber is calculated from
  $k=-\partial\tilde{\omega}/\partial a$. 
  The grey line is the expected wavenumber,
  Eqn.\ (\ref{k}), with $\varepsilon=1$. \vspace*{2in}
}
\end{figure}

\newpage
\begin{figure}
\epsscale{1.0}
\plotone{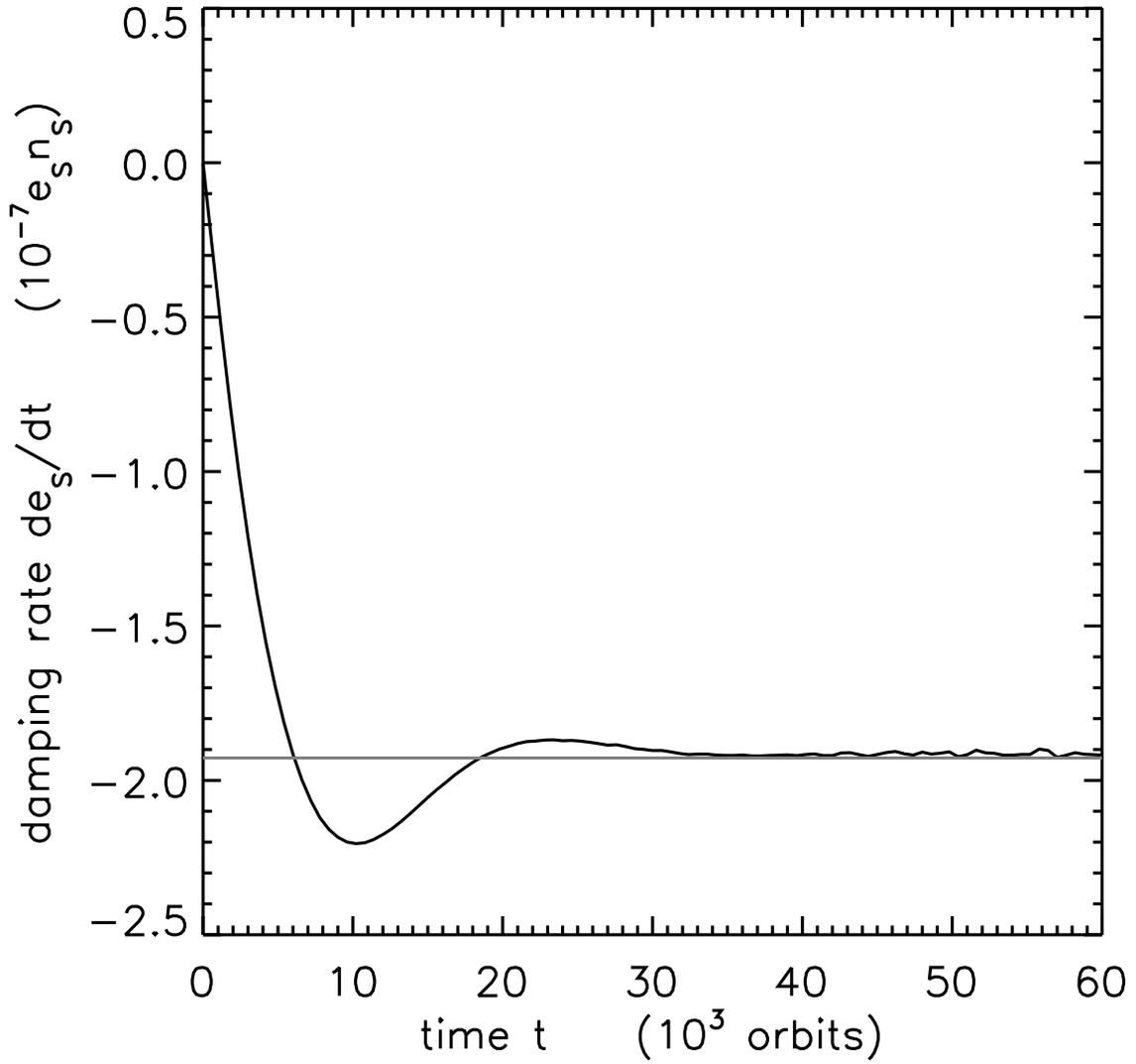}
\figcaption{
  \label{edot_fig}
  The rate at which the satellite launching the wave in
  Fig.\ \ref{ewave} has its eccentricity damped,
  $\dot{e}_s$, plotted versus time t in units of $10^3$ orbit periods. 
  The solid gray curve is the expected rate, 
  Eqn.\ (\ref{edamping2}).
}
\end{figure}

\newpage
\begin{figure}
\epsscale{1.0}
\plotone{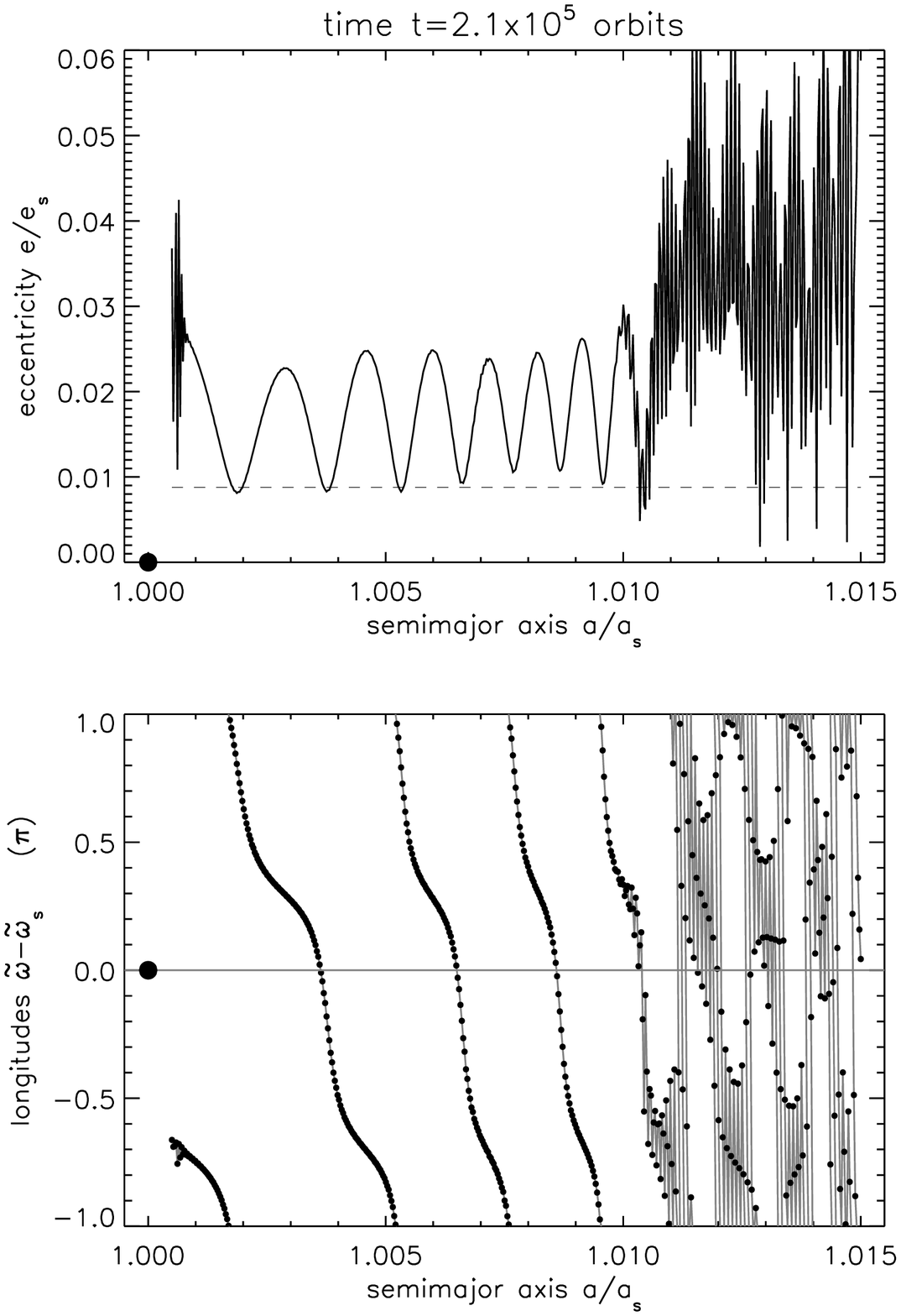}
\end{figure}

\newpage
\begin{figure}
\figcaption{
  \label{ew_fig}
  The simulation described in Section \ref{outbound waves} and 
  Figs.\ \ref{ewave}--\ref{edot_fig} is extended to
  time $t=2.1\times10^5$ orbits. During this time, the outbound
  long wave has since reflected at the simulated disk's outer edge
  at $a/a_s=1.015$ as a superposition of inbound long and
  short density waves.
  The long-wavelength undulations seen in $e(a)$ thus represent
  the superposition of the outbound and inbound long waves, while
  the high-frequency variations in $e(a)$ are due to the slower moving
  short waves. See Section \ref{reflect} for details.
}
\end{figure}

\end{document}